\documentclass{aa}

\usepackage{mathrsfs}
\usepackage{hyperref}

\usepackage{subcaption}
\usepackage{txfonts}
\usepackage{nicefrac}
\usepackage{graphicx}
\begin{document}

   \title{Resolving nearby dust clouds}

   \author{R. H. Leike\inst{1}\inst{2}
          \and
          M. Glatzle \inst{1}\inst{3}
          \and
          T. A. En\ss lin \inst{1}\inst{2}
          }

   \institute{Max Planck Institute for Astrophysics, Karl-Schwarzschildstra\ss e 1, 85748 Garching, Germany
         \and
         Ludwig-Maximilians-Universit\"at, Geschwister-Scholl Platz 1, 80539 Munich, Germany
         \and
         Physik-Department, Technische Universit\"at M\"unchen, James-Franck-Str. 1, 85748 Garching, Germany
       }

   \date{Received XXXX, accepted XXXX}

% \abstract{}{}{}{}{} 
% 5 {} token are mandatory
 
  \abstract
  {}
   {Mapping the interstellar medium in 3D provides a wealth of insights into its inner working. The Milky Way is the only galaxy for which detailed 3D mapping can be achieved in principle. In this paper, we reconstruct the dust density in and around the local super-bubble.}
    {The combined data from surveys such as Gaia, 2MASS, PANSTARRS, and ALLWISE provide the necessary information to make detailed maps of the interstellar medium in our surrounding. To this end, we used variational inference and Gaussian processes to model the dust extinction density, exploiting its intrinsic correlations.}
   {We reconstructed a highly resolved dust map, showing the nearest dust clouds at a distance of up to $400\,\text{pc}$ with a resolution of $1\,\text{pc}$.}
   {Our reconstruction provides insights into the structure of the interstellar medium. We compute summary statistics of the spectral index and the 1-point function of the logarithmic dust extinction density, which may constrain simulations of the interstellar medium that achieve a similar resolution.}


   \maketitle
%
%________________________________________________________________

\section{Introduction}
\label{sec:introduction}

Although dust contributes only a small fraction in terms of mass, it is an important constituent of the interstellar medium (ISM) that is observable in many wavebands of the electromagnetic spectrum. %
Dust efficiently absorbs and scatters ultra-violet and visible range photons, obscuring large parts of the Galaxy and hiding star forming regions at these wavelengths. %
The dust absorbed energy is re-emitted in the infrared to microwave bands, offering a diagnostic for physical conditions of the ISM. %
The microwave emission of dust is a significant foreground to the cosmic microwave background (CMB). %

Dust plays a role in many processes that drive galactic evolution. %
Grain surfaces can adsorb material from interstellar gas and act as catalytic sites for chemical reactions. %
Stars, including the most massive ones, are observed to form from dusty molecular clouds. %
Thermal emission from dust grains can be an important cooling channel for these clouds and grains can drive their chemistry, suggesting that dust plays an important role in regulating the star formation process. %
Photons absorbed by dust can convey radiation pressure to interstellar matter, or, if they are energetic enough, eject electrons, contributing to the heating of interstellar gas. %

Finally, the distribution of dust can be used as a tracer of other quantities. %
A significant portion of the observed Galactic gamma rays in the GeV-range originates in dense clouds, where it is produced by hadronic interactions of cosmic rays with gas. %
This can be seen, for example, in the morphology of cosmic rays with hadronic spectrum from FERMI \citep{selig2015denoised}. %
Dust can be used to trace these dense clouds and identify gamma-ray production sites. %
Another example is the magnetic field structure of the Galaxy, which is imprinted in the dust density, as dust filaments tend to be aligned to the line-of-sight magnetic field \citep{panopoulou2016magnetic}. %
Dust also reveals the large scale dynamics and structure of the Galaxy, as the gravitational and differential rotation imprints on the filaments of dust. %

Studying how dust is distributed in the Galaxy can not only provide an understanding of its contents and structure, but also into its inner workings, and aid in the interpretation of observations in dust-affected wavebands. %
Most 3D mapping efforts so far have aimed to reconstruct the distribution of dust in our Galaxy on large scales. %
This is interesting as it reveals the structure of our Galaxy, such as its spiral arms. %
Some notable recent contribution in this direction was provided by \citet{green20193d}, who mapped three quarters of the sky using Gaia, 2MASS, and PANSTARRS data using importance sampling on a gridded parameter space and by assuming a Gaussian process prior. %
\citet{lallement2019gaia} reconstructed a map extending out to $3\,\text{kpc}$ with a $25\,\text{pc}$ resolution based on Gaia and 2MASS data with Gaussian process regression. %
\citet{chen2018three} reconstructed a map extending out to $6\,\text{kpc}$ with a $0.2\,\text{kpc}$ radial resolution based on Gaia, 2MASS, and WISE data with random forest regression. %

This paper can be regarded as a follow-up to \citet{leike2019charting}. %
Some derivations are kept short here, and we advise \citet{leike2019charting} as a co-read for the statistically inclined reader. %
We focus on reconstructing only the nearby dust clouds, within $\sim 400\,\text{pc}$. %
While this prohibits revealing spiral arms, it enables us to achieve higher resolution. %
This way, we hope to be able to constrain simulations of the ISM, which achieve a similar resolution. %
The map might also prove relevant for foreground corrections to the CMB, especially for CMB polarization studies. %
It was shown that most of the Galactic infrared polarization at high latitudes ($|b|>60$) comes from close-by regions around $200$-$300$~pc \citep{skalidis2019local}. %
Correction maps have so far been based on infrared observations, and could be biased through different starlight illumination or differing dust temperatures. %

%______________________________________________________________

\section{Data}
\label{sec:data}

For our 3D reconstruction, we used combined observational data of Gaia DR2, ALLWISE, PANSTARRS, and 2MASS. %
These datasets were combined and processed to yield one consistent catalog with stellar parameters by \citet{anders2019photo}. %
We used these high-level preprocessed data for our reconstruction. %
Table\,\ref{table:used-data} contains a summary of the columns we extracted from this dataset. %
We further selected only sources that are inside an $800\,\text{pc}\times800\,\text{pc}\times600\,\text{pc}$ cube centered on the Sun. %
To determine whether a source is inside this cube, we used their $84\%$ distance quantile $\text{dist}_{84}$. %
We assumed a Gaussian error on the parallax, with mean $m_\omega$ and standard deviation computed from the distance quantiles as
\begin{align}
    m_\omega &= \frac{1}{2}\left(1/\text{dist}_{16}+1/\text{dist}_{84}\right)\label{eq:parallax-mean}\\
    \sigma_\omega &= \frac{1}{2}\left(1/\text{dist}_{16}-1/\text{dist}_{84}\right)\label{eq:parallax-std} \,.
\end{align}
Furthermore, we applied the following selection criteria;
\begin{align}
    \text{SH\_OUTFLAG} &= \text{00000},\\
    \text{SH\_GAIAFLAG} &= \text{000},\\
    \text{ph} &\in \text{Table } \ref{table:good-photoflags} ,\\
    \nicefrac{\sigma_\omega}{m_\omega} &< 0.3,\\
    \text{av}_{05} &\neq \text{av}_{16} \,.
\end{align}
In other words, we selected only stars that have clean starhorse pipeline flags, a clean Gaia flag, a specific photo flag, and sufficiently small parallax error. %
We required the constraint on the photo flag, because we only derived the noise statistic for stars with this flag. %
For details, see Sect.\,\ref{sec:noise-statistic}. %
Additionally, we excluded stars for which the $5\%$ V-band extinction quantile is equal to the $16\%$ quantile, as this suggests that the pipeline had difficulties for these sources. %

\begin{figure}
        \includegraphics[width=0.5\textwidth]{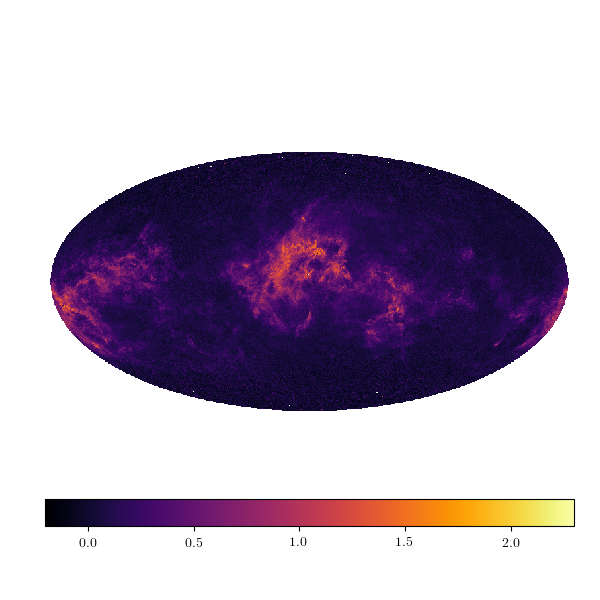}
    \caption{A Mollweide projection of the $G$-band extinction optical depth $a$ to all sources in the used dataset. For this healpix nside 128 plot, we average the data sources that are in the same pixel, using the inverse noise dispersion as weights. Pixels with no data appear in white.} %
        \label{fig:our-data}
\end{figure}

These criteria result in the selection of a total of 5\,096\,642 sources. %
Figure\,\ref{fig:our-data} shows an inverse-noise weighted average of our data projected onto the sky. %
To consistently combine the information of many data points, it is crucial to know the likelihood of a data point given the true amount of extinction for that source. %
We call this likelihood of one data point given its true extinction the noise statistic to distinguish it from the likelihood of the whole dataset given the true 3D dust extinction distribution, which contains additional operations (see Sect.\,\ref{sec:likelihood} for details). %
Unfortunately, \cite{anders2019photo} did not publish a noise statistic for their dataset, and a noise statistic is not readily derivable from posterior quantiles. %
This is because posterior quantiles $a$ give very limited information on the distribution $P(a^*|a)$ of the true extinction $a^*$, while a full noise statistic would be given by $P(a|a^*)$. %
In particular, there is no natural way to derive an analytic form of $P(a^*|a)$, inhibiting the calculation of $P(a|a^*)$. %
We thus chose a different approach to infer the noise statistic, which we describe in Sect\,\ref{sec:noise-statistic}.

\begin{table*}
\centering
\footnotesize
\begin{tabular}{ccc}
    name in \citet{anders2019photo} & our notation & explanation\\
\hline
    dist16 & $\text{dist}_{16}$ & $16\%$ distance quantile\\
    dist50 & $\text{dist}_{50}$ & $50\%$ distance quantile\\
    dist84 & $\text{dist}_{84}$ & $84\%$ distance quantile\\
    ag50 & $a$ & $50\%$ $G$-band extinction quantile\\
    SH\_PHOTOFLAG & ph & photo bands used for data point\\
    SH\_GAIAFLAG & SH\_GAIAFLAG & output flag of Gaia\\
    SH\_OUTFLAG & SH\_OUTFLAG & output flag of the starhorse pipeline\\
\end{tabular}
\normalsize
    \caption{Data columns extracted from \citet{anders2019photo}}
\label{table:used-data}
\end{table*}
\section{Likelihood}
\label{sec:likelihood}

\subsection{Response}
\label{sec:response}

If we have the true 3D extinction density $s(x)$, we can compute the extinction $a^*_i$ for each source $i$ by computing the line integral $R_i$ %
\begin{align}
    a^*_i = R_i^{\omega*}(s) = \int_0^{\frac{1}{\omega^*_i}} s(r\theta_i) \text{d}r \,,
\end{align}
where $\theta_i$ is the position of the $i$-th source projected onto the unit sphere and $\omega^*_i$ is the true parallax of the source. %
The true parallax $\omega^*_i$ is assumed to be Gaussian distributed, with error and mean computed from the $16\%$ and $84\%$ percentiles of the starhorse dataset according to Eqs.\,(\ref{eq:parallax-mean}) and (\ref{eq:parallax-std}): %
\begin{align}
    \omega^*_i \curvearrowleft \mathscr{G}(\omega^*_i|\frac{1}{2}(\omega_{84,i}+\omega_{16,i}), \frac{1}{4}(\omega_{84,i}-\omega_{16,i})^2)\,.\label{eq:true-parallax}
\end{align}
Given this uncertainty of the true source distance, we can compute the expected extinction density for the source $i$ as a weighted line integral $R_i$
\begin{align}
    \left<a^*_i\right>_{P(\omega^*_i|\omega_{16,i},\omega_{84,i})} &= R_i(s) = \left<R^{\omega^*}_i(s)\right>_{P(\omega^*_i|\omega_{16,i},\omega_{84,i})}\nonumber\\ 
    &=\int_0^{\frac{1}{\omega^*_i}} s(r\theta_i)(1-\text{cdf}(r|\omega_{16,i},\omega_{84,i})) \text{d}r\ ,\label{eq:response}
\end{align}
where $\text{cdf}$ denotes the cumulative density function of Eq.\,(\ref{eq:true-parallax}) with $r=(\omega^*)^{-1}$. %
We computed the line integral of Eq.\,(\ref{eq:response}) on the fly for every step, using a parallelized fortran code\footnote{\url{https://gitlab.mpcdf.mpg.de/mglatzle/gda_futils}}. %

The uncertainty of the true position of the source introduces a source-dependent supplementary noise contribution $\widehat{\sigma}^2_i$. %
This uncertainty arises due to the uncertainty of the true source distance, which introduces uncertainty on the line-of-sight extinction even when given the true extinction density $s$. %
The standard deviation of this supplementary noise contribution can be computed as %
\begin{align}
    \widehat{\sigma}^2_{i} &= {\text{Var}\left[P\left(a^*_i|\omega_{16,i},\omega_{84,i}, s\right)\right]}\nonumber\\
    &= {\text{Var}\left[\int\text{d}\omega^*_i\,P\left(a_i^*,\omega^*_i|\omega_{16,i},\omega_{84,i}, s\right)\right]}\nonumber\\
    &= {\text{Var}\left[\int\text{d}\omega^*_i\,P\left(a_i^*|s, \omega^*_i\right)P\left(\omega^*_i|\omega_{16,i},\omega_{84,i},s\right)\right]}\nonumber\\
    &\leq {\text{Var}\left[\int\text{d}\omega^*_i\,P\left(a_i^*|s, \omega^*_i\right)P\left(\omega^*_i|\omega_{16,i},\omega_{84,i}\right)\right]}\nonumber\\
    &={\text{Var}\left[\int\text{d}\omega^*_i\,\delta\left(a_i^*-R_i^{\omega^*}(s)\right)P\left(\omega^*_i|\omega_{16,i},\omega_{84,i}\right)\right]}\,.\label{eq:supplementary-noise}
\end{align}
The last inequality holds, as $P(\omega^*_i|\omega_{16,i},\omega_{84,i})$ has strictly more variance than $P(\omega^*_i|\omega_{16,i},\omega_{84,i},s)$. %
We sampled this additional noise contribution before every step of our algorithm. %
We did this by drawing $M=20$ samples $j$ of parallaxes $\omega_i^j$ according to the statistic given by Eq.\,(\ref{eq:true-parallax}). %
We then computed 
\begin{align}
    \widehat{\sigma}^2_{i} = \frac{1}{M}\sum_j R^{\omega^j_i}(s)
\end{align}
as the sample variance of the extinction estimate using the samples $j$ and the current reconstructed dust extinction density $s$. %
This error correction was not done in \citet{leike2019charting}. %
However, for this paper, the smaller data uncertainty and slightly higher parallax error of the sources raises the importance of computing this error correction, while the use of the new code for the response enables its calculation. %
%Here $\mathbb{1}_{[a,b]}$ denotes the indicator function on the interval $[a,b]$, defined via %
%begin{align}
%   \mathbb{1}_{[a,b]}(x) = 
%   \begin{cases}
%       1 & x \in [a,b]\\
%       0 & \text{otherwise}
%   \end{cases}
%end{align}

\subsection{Noise statistic}
\label{sec:noise-statistic}

The noise statistic specifies how probable an observed $G$-band extinction value is, given that one would know the true amount of $G$-band extinction for that source. %
Since there is no detailed noise statistic published for the dataset we used, we had to construct it. %
To do this, we looked at regions of the sky where there is no significant amount of dust expected. %
These regions were identified by using the Planck dust map \citep{akrami2018planck}, more specifically the dust map from the COMMANDER pipeline of the 2014 Planck data release. %
Here, regions with less than $\text{exp}(2)\nicefrac{\mu K}{rJ}$ were taken to be dustless. %
This criterion selects 606 pixels of the healpix nside 256 dust map, corresponding to $0.077\%$ of the sky. %
For every SH\_PHOTOFLAG for which we have more than 100 values in these dustless regions, we calculated the mean $m_\text{ph}$ and standard deviation $\sigma_\text{ph}$ of all $G$-band extinctions. %
Using these values, we define the probability to measure an extinction $a$ given the true extinction $a^*$ as %
\begin{align}
    P(a|a^*, \text{SH\_PHOTOFLAG}=\text{ph}) = \mathscr{G}(a|a^*+m_\text{ph}, \sigma_\text{ph}^2)\label{eq:data-distribution} \ .
\end{align}
Table \ref{table:good-photoflags} shows our used means $m_\text{ph}$ and standard deviations $\sigma_\text{ph}$ for all used photo flags $\text{ph}$. %
As can be seen by investigating Table \ref{table:good-photoflags}, the mean values deviate strongly from zero, and correcting the zero point is vital to our reconstruction. %
\begin{table}
\centering
\footnotesize
\begin{tabular}{ccc}
    \text{ph}=SH\_PHOTOFLAG & mean $m_\text{ph}$ & standard deviation $\sigma_\text{ph}$\\
\hline
GBPRP & 0.493  & 0.439  \\
GBPRPJHKs & 0.131  & 0.259 \\
GBPRPJHKs\#W1W2 & 0.315  & 0.538 \\
GBPRPJHKsW1W2 & 0.116 & 0.232 \\
GBPRPgrizyJHKs & 0.223 & 0.209 \\
GBPRPgrizyJHKsW1W2 & 0.156 & 0.172 \\
GBPRPiJHKsW1W2 & 0.101 & 0.219 \\
GBPRPiyJHKsW1W2 & 0.165 & 0.234 \\
\end{tabular}
\normalsize
    \caption{SH\_PHOTOFLAG values and the corresponding mean and standard deviations for sources in dustless regions. Regions are considered as dustless if the Planck dust map shows weaker emission than $\text{exp}(2)\nicefrac{\mu K}{rJ}$.}
\label{table:good-photoflags}
\end{table}
We note that because we fix the noise statistic for an actual extinction value of zero, the reconstruction might be biased for high extinction values. %
We discuss some biases that could be attributed to this effect in Sect.\,
\ref{sec:comparison}.

\section{Prior}
\label{sec:prior}

We folded our physical knowledge into the prior of the dust extinction density. %
We chose the exact same prior model as in \cite{leike2019charting}. %
We assumed the extinction density $s$ to be positive and spatially correlated. %
This can be enforced by assuming a log-normal Gaussian process prior %
\begin{align}
    s_x &= \rho_0\,\text{exp}(\tau_x) \,,\\
    \tau &\curvearrowleft \mathscr{G}(\tau|0,T) \,,
\end{align}
where $\rho_0$ is the prior median extinction density and $T$ is the correlation kernel of the Gaussian process $\tau$. %
The prior median extinction density is a hyperparameter of our model, and we chose $\rho_0=\nicefrac{1}{1000}\text{pc}^{-1}$. %
We inferred the kernel $T$ during our reconstruction. %
This can be achieved by rewriting $s$ in terms of a generative model %
\begin{align}
    s_x = \rho_0 \,\text{exp}(\mathbb{F}\sqrt{T_k(\xi_T)}\xi_k) \,,
\end{align}
where all $\xi$ are a-priori standard normal distributed, and $T_k(\xi_T)$ is a nonparametric model for the Fourier-transformed correlation kernel $T_k$, also called the spatial correlation power spectrum. %
One should note that this model is degenerate: any change in $T_k$ can be absorbed into $\xi_k$ instead, as only the product of these two fields enters the overall dust extinction density $s$. %
Because of this property, the reconstructed power spectrum $T_k$ does not have to be the empirical power spectrum of $s_x$, which can be calculated by Fourier transforming and binning. %
To avoid misunderstandings and artifacts from the degenerate model, we mainly report the empirical power spectrum in this paper, which is computed from posterior samples of $s_x$. %
We now focus on our model for the power spectrum $T_k$. This model assumes the spatial correlation power spectrum to be a preferentially falling power law, but allows for arbitrary deviations. %
It can be written as %
\begin{align}
    \sqrt{T_k(\xi_T)} &= \text{Exp}^*\text{Exp}\big[(m_s+\sigma_s\xi_s)\text{ln}(k)+m_0+\sigma_0\xi_0 \nonumber\\
    &\quad+ \mathbb{F}_{\text{ln}(k)t}\text{sym}(\nicefrac{A}{\left(1+\left(\nicefrac{t}{t_0}\right)^2\right)}\xi_\phi(t))\big]\ ,
\end{align} %
where the first part describes a linear function on log-log scale, for instance a power law; and the second part describes the nonparametric deviations that are assumed to be differentiable on log-log scale. %
The operation $\text{Exp}^*$ denotes the exponentiation of the coordinate system. %
More explicitly, $\mathbb{F}_{\text{ln}(k)t}$ is Fourier transformation on log-log scale, and the function sym is defined as %
\begin{align}
     f: [0,2b]\rightarrow \mathbb{R}\\
    \text{sym}(f)(x) = \left(f(x)-f(2b-x)\right)\big|_{[0,b]}\ ,
\end{align}
where $f\big|_M$ denotes the restriction of the domain of the function $f$ to $M$. %
The function sym is required to deal with the periodic boundary conditions introduced by the Fourier transform. %
Details can be found in the appendix of \citet{arras2019unified}. %
The hyperparameters of the model are $(A, t_0, m_s, \sigma_s, m_0, \sigma_0),$ which we chose to be $(11, 0.2, -4, 1, -14, 3)$ in complete analogy to \citet{leike2019charting}. %

\section{Algorithm}
\label{sec:algorithm}

We combined the prior and the likelihood into one generative model of the data. %
We computed approximate posterior samples using metric Gaussian variational inference (MGVI) \citep{knollmuller2019metric}. %
This variational approach alternates between drawing samples around the current estimate for the latent parameters and optimizing the current estimate using the average gradient of the samples. %
The final set of samples was used to derive an uncertainty estimate on all our maps, as well as on all derived quantities. %

For further parallelization, we split the problem into the eight octants. %
Each octant measures $410\text{pc}\times410\text{pc}\times310\text{pc}$,
meaning that they overlap for $20$pc. %

We hereby used a threefold parallelization scheme, parallelizing by octants, parallelizing by samples, and a parallelized response. %
The latter two parallelizations were enabled by our new fortran implementation, which computes the arising line integrals (Eq.\,(\ref{eq:response})) on the fly. %
This is in contrast to our previous paper \cite{leike2019charting}, where we computed the line integral using sparse matrices. %
Computing the response on the fly takes approximately the same amount of time, but does not have any additional memory requirements, and therefore allows for parallelization and a larger reconstruction. %

The total number of degrees of freedom is $\approx 417 \text{ million}$, exceeding those of our previous map by a factor of 30. %
The total computation time was about two weeks of wall clock time, or about $0.5$ million CPUh on $1920$ cores. %

The final samples of the independently reconstructed octants are combined into the full reconstruction using a differentiable variance-preserving interpolation scheme. %
The details are described in Appendix\,\ref{sec:interpolation-scheme}.
One noteworthy point is that we cut away the outer 30pc due to artifacts from periodic boundary conditions, resulting in a final map volume of $740\,\text{pc}\times740\,\text{pc}\times540\,\text{pc}$. %

\section{Results and discussion}
\label{sec:results}
\subsection{Results}
% Here we discuss what we found out

\begin{figure*}[p]
        \centering
        \begin{subfigure}[t]{.46\textwidth}
                \includegraphics[trim={0.8cm .2cm .5cm .5cm}, clip, width=.9\textwidth]{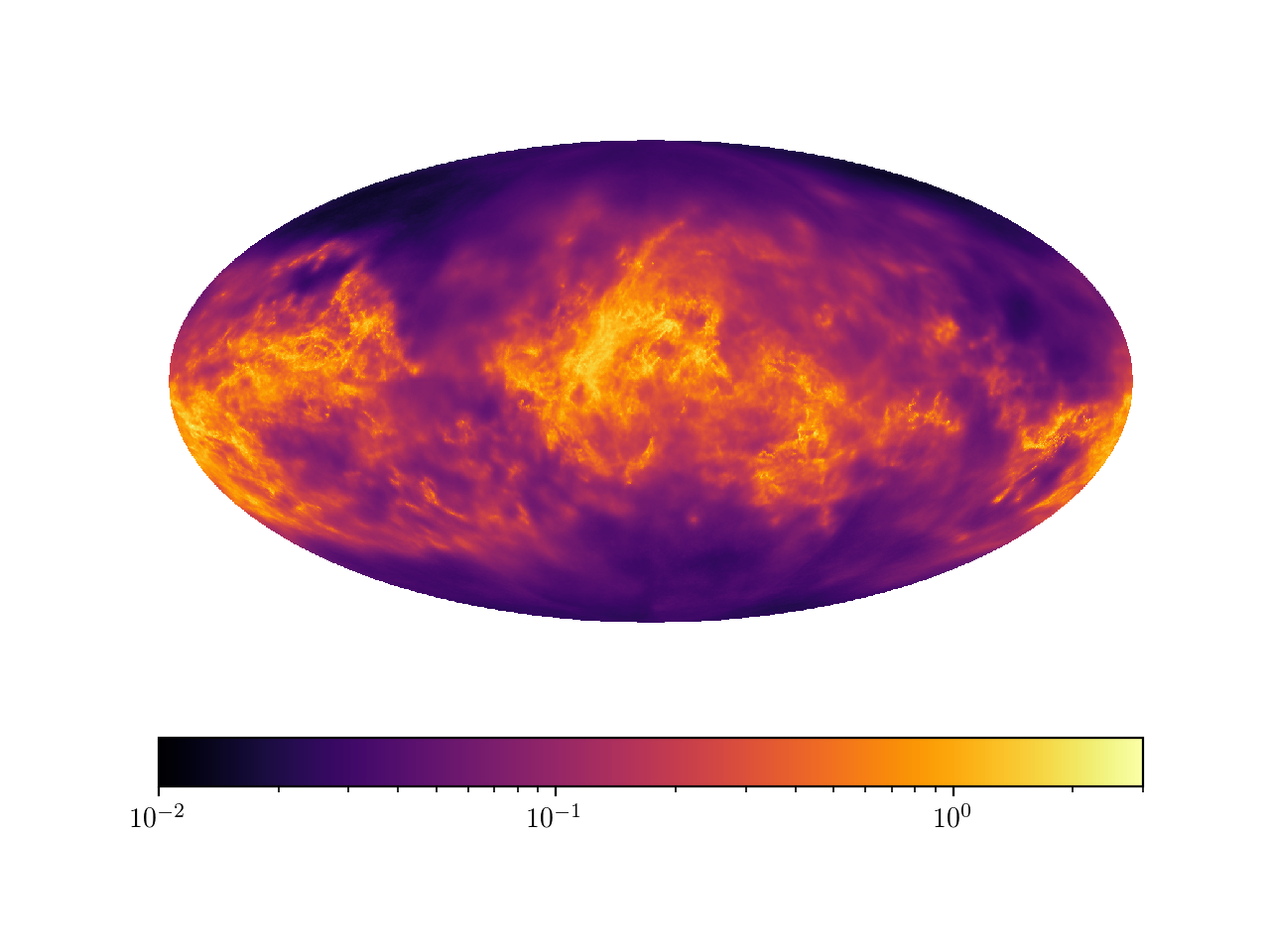}
        \caption{
        }
        \end{subfigure}
        ~
        \begin{subfigure}[t]{.46\textwidth}
                \includegraphics[trim={0.8cm .2cm .5cm .5cm}, clip, width=.9\textwidth]{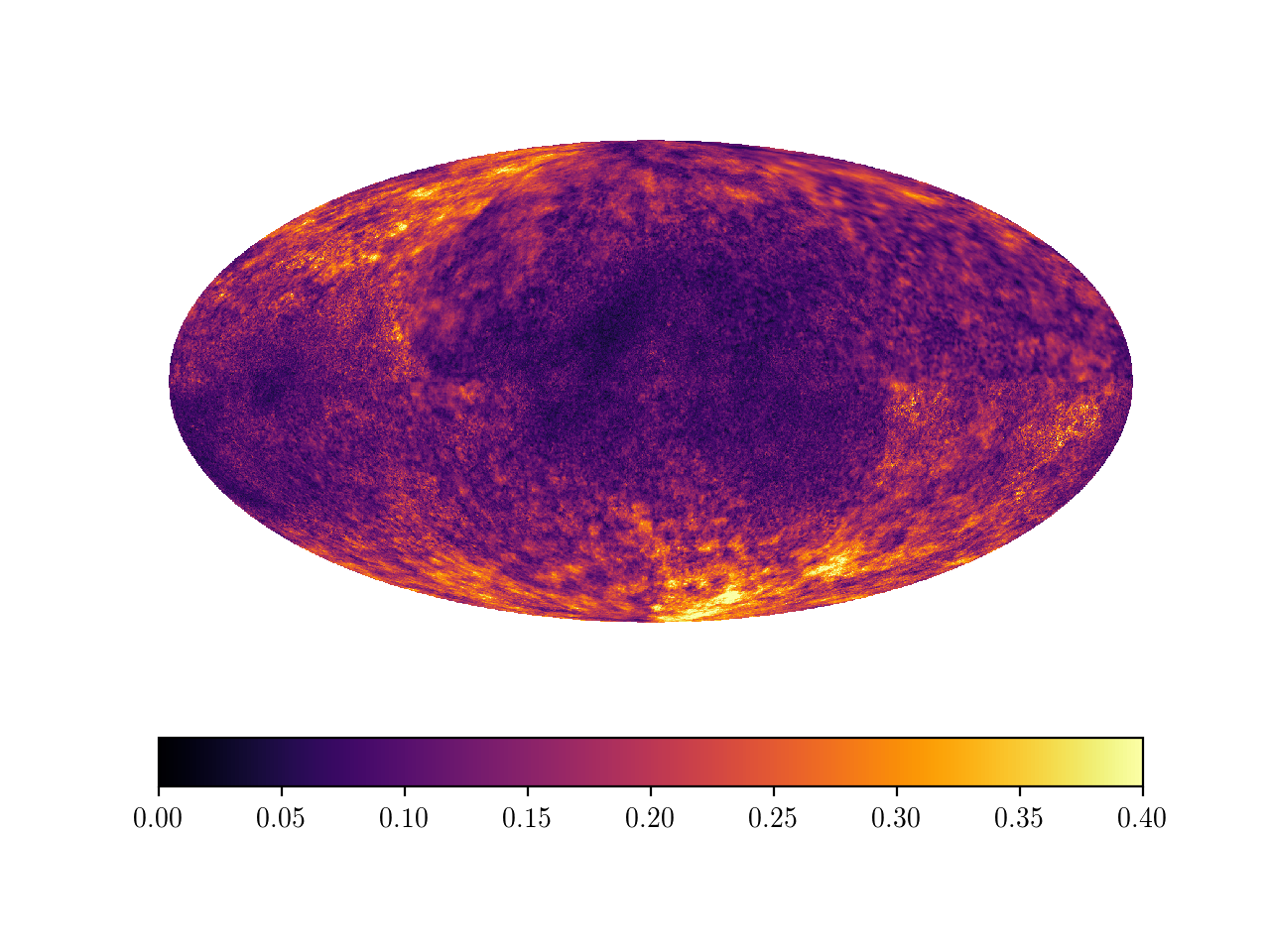}
                \caption{
}
        \end{subfigure}
        \\
        \begin{subfigure}[t]{.46\textwidth}
                \includegraphics[trim={0.8cm .2cm 2.0cm 1cm}, clip, width=.95\textwidth]{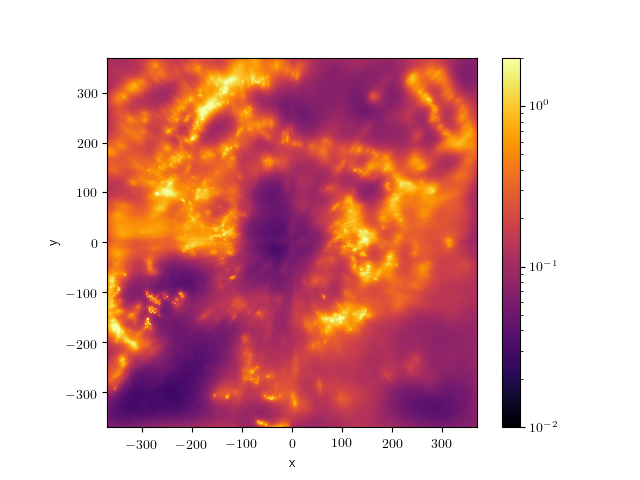}
        \caption{
        }
        \end{subfigure}
        ~
        \begin{subfigure}[t]{.46\textwidth}
        \includegraphics[trim={0.8cm .2cm 2.0cm 1cm}, clip, width=.95\textwidth]{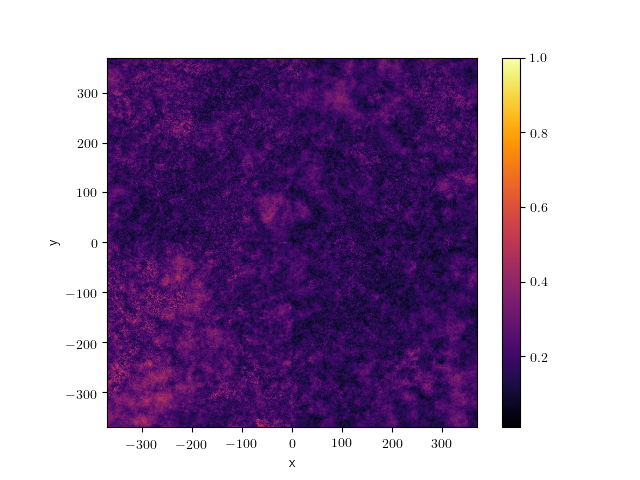}
        \caption{
                }
        \end{subfigure}
        \\
        \begin{subfigure}[t]{.46\textwidth}
        \includegraphics[trim={0.8cm .2cm 2.0cm 1cm}, clip, width=.95\textwidth]{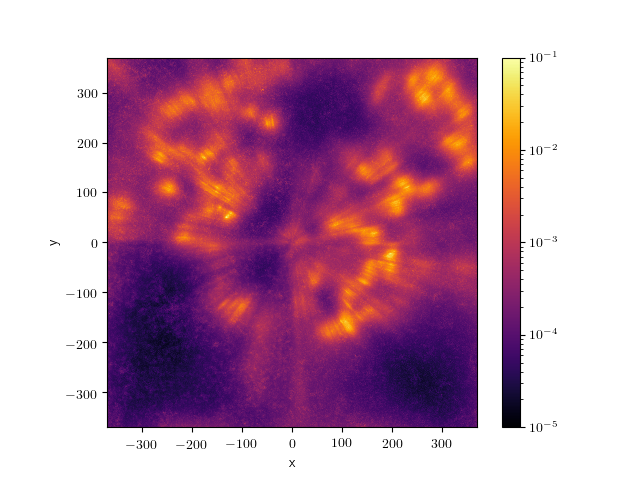}
        \caption{
        }
        \end{subfigure}
        ~
        \begin{subfigure}[t]{.46\textwidth}
                \includegraphics[trim={0.8cm .2cm 2.0cm 1cm}, clip, width=.95\textwidth]{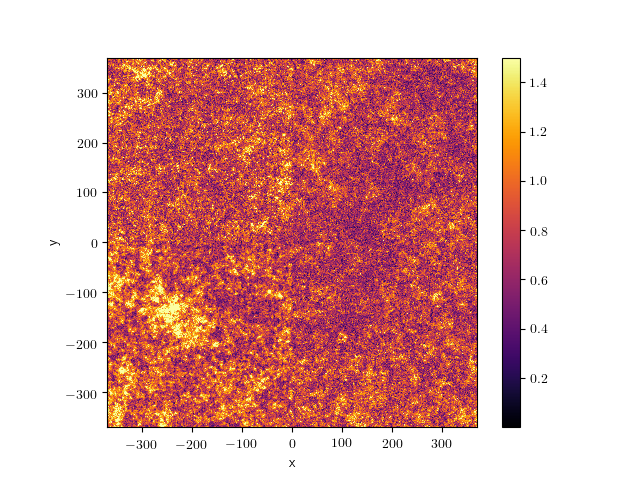}
        \caption{
        }
        \end{subfigure}
    \caption{\label{fig:result}
    Result of our 3D dust reconstruction.
    The first column shows dust extinction, the second shows the relative error.
    The first row shows the integrated extinction in e-folds in a Mollweide projection of the whole reconstructed box of $740\,\text{pc}\times740\,\text{pc}\times540\,\text{pc}$. The second row also shows integrated extinction in e-folds in the same box, but integrated normally to the Galactic plane instead of radially.
    The third row shows differential extinction in e-folds per parsec in a slice along the Galactic plane.
        }                       %
\end{figure*}

We were able to reconstruct the nearby dust clouds. %
Figure\,\ref{fig:result} shows various maps produced from our result and their relative uncertainty. %
The maps show tendrils and filaments of dust on scales as small as $2\text{pc}$ up to scales of several hundred parsecs, at which they become disconnected. %

\begin{figure}
                \includegraphics[trim={0.8cm .2cm .5cm 0.5cm}, clip, width=.45\textwidth]{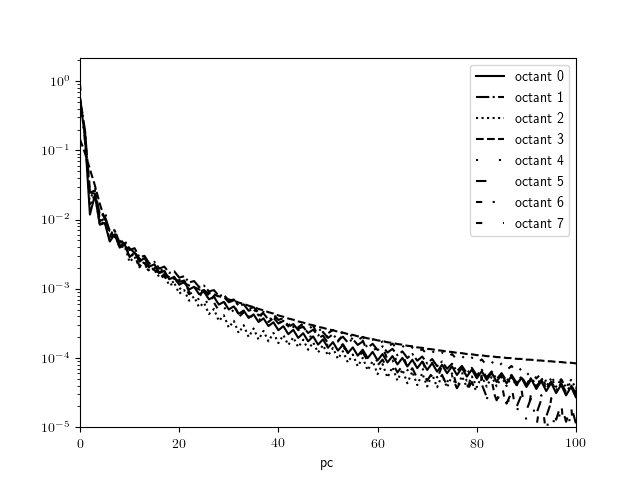}
        \caption{\label{fig:kernels}
        Reconstructed correlation kernels for the different octants.
        We note that the logarithmic dust extinction in our model is the result of an a-priori normal distributed field that is folded with these kernels, dependent on the octant.
    The octants are arranged such that octant $i= 4b_2 + 2b_1 + b_0$ (for $b_i\in\{0,1\}$) extends in positive $x$-direction if and only if $b_0=0$, in positive $y$-direction if and only if $b_1=0$ and in positive $z$-direction if and only if $b_2=0$. %
    We note that all kernels fall to about $10\%$ in the first $2\,\text{pc}$.
}                               %
\end{figure}

All octants inferred similar logarithmic convolution kernels, as can be seen in Fig. \ref{fig:kernels}. %
These correlation kernels were computed by taking a slice out of the reconstructed Fourier-transformed square root power spectrum. %

\begin{figure}
                \includegraphics[trim={0.4cm .2cm .5cm 0.5cm}, clip, width=.45\textwidth]{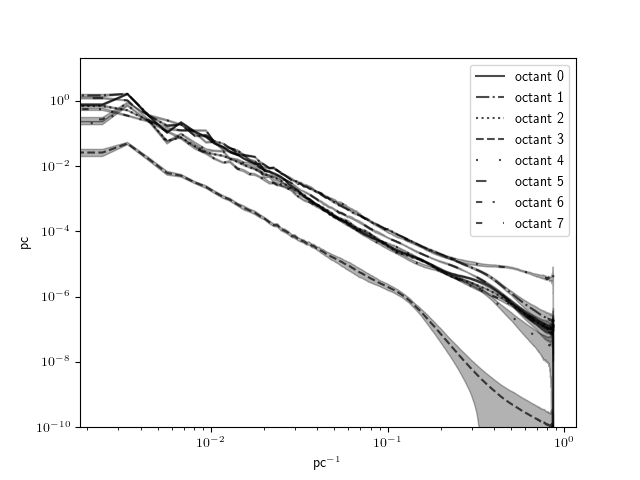}
        \caption{\label{fig:octant-power}
        Empirical power spectra of the dust extinction density of the eight octants. %
    The octants are arranged such that octant $i= 4b_2 + 2b_1 + b_0$ (for $b_i\in\{0,1\}$) extends in positive $x$-direction if and only if $b_0=0$, in positive $y$-direction if and only if $b_1=0$ and in positive $z$-direction if and only if $b_2=0$. %
}                               %
\end{figure}

A comparison of the empirical power spectra of the eight different octants can be found in Fig.\,\ref{fig:octant-power}. %
Most octants have very similar power spectra, only octant three deviates strongly. %
This octant, located at $180<l\leq270$ and $b>0$ (disregarding the overlap), is strongly devoid of dust, explaining the significantly lower power spectrum. %

For the power of the full-volume extinction density, we find a power law with spectral index of $2.52\pm 0.015$ at scales from $2$pc to $100$pc. %
For the logarithmic power, we report a spectral index of $2.82\pm0.022$ at scales from $2.3$pc to $125$pc. %

\begin{figure*}
    \begin{subfigure}[t]{.46\textwidth}
                \includegraphics[trim={0.8cm .2cm .5cm 0.5cm}, clip, width=.95\textwidth]{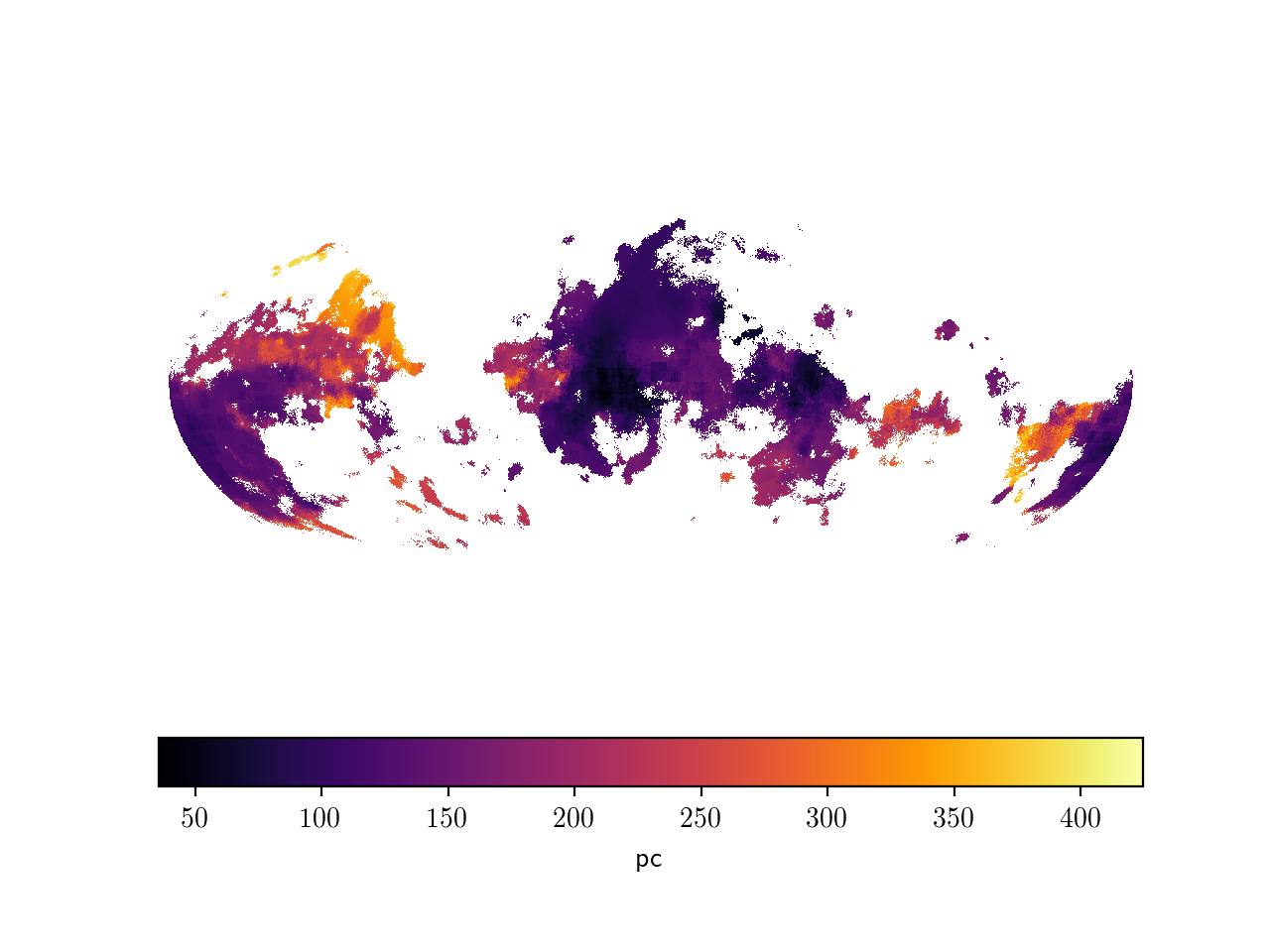}
    \end{subfigure}
    \begin{subfigure}[t]{.46\textwidth}
                \includegraphics[trim={0.8cm .2cm .5cm 0.5cm}, clip, width=.95\textwidth]{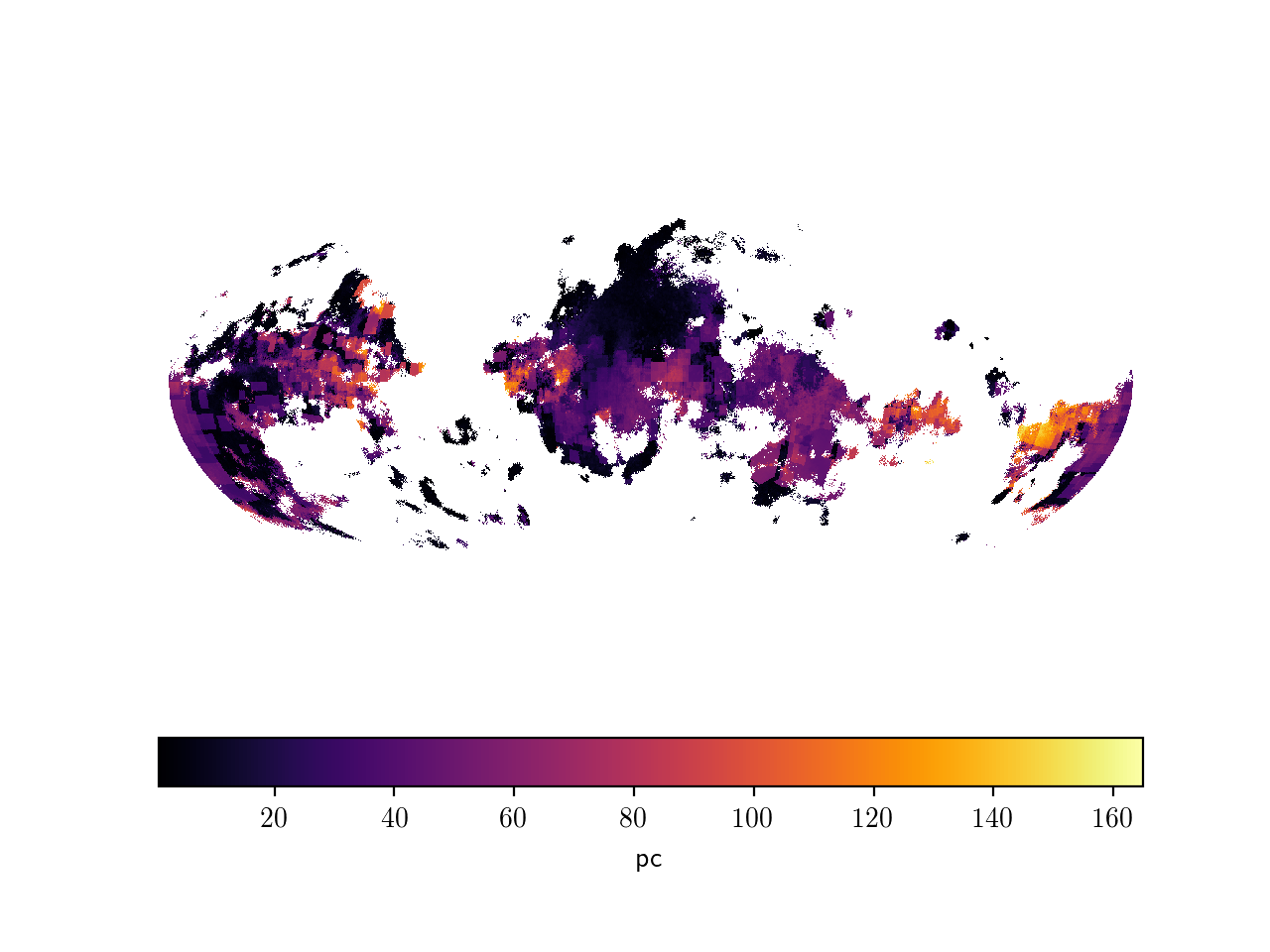}
    \end{subfigure}
        \caption{\label{fig:dist-to-nearest}
        A Mollweide projection showing the distance to the first voxel of our reconstruction that exceeds an extinction estimate of $0.005$ e-folds per parsec (left side) and corresponding uncertainty map (right side).
        Directions for which the threshold is never reached appear in white.
}                               %
\end{figure*}

\begin{figure*}
    \begin{subfigure}[t]{.46\textwidth}
                \includegraphics[trim={0.8cm .2cm .5cm 0.5cm}, clip, width=.95\textwidth]{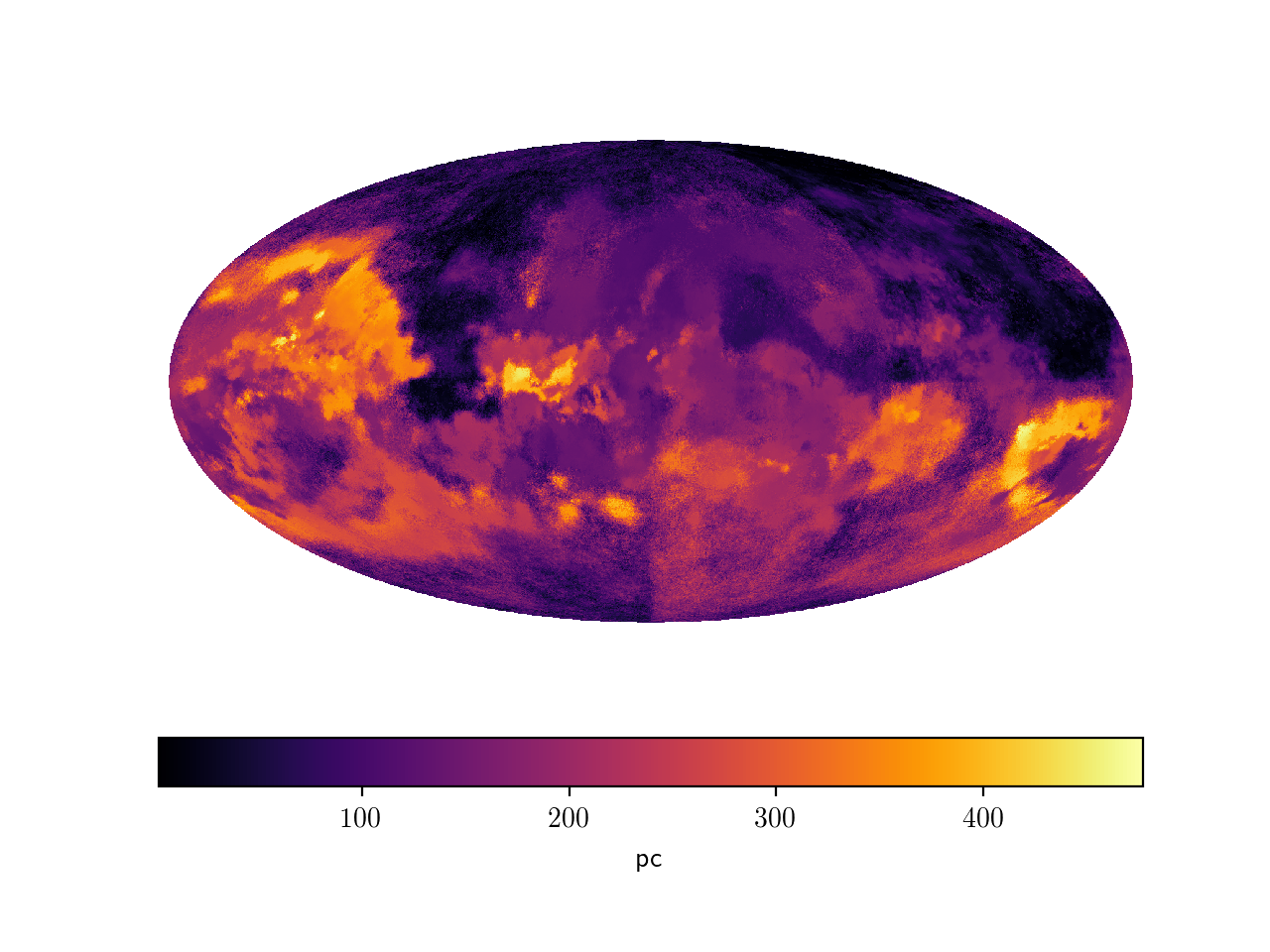}
    \end{subfigure}
    \begin{subfigure}[t]{.46\textwidth}
                \includegraphics[trim={0.8cm .2cm .5cm 0.5cm}, clip, width=.95\textwidth]{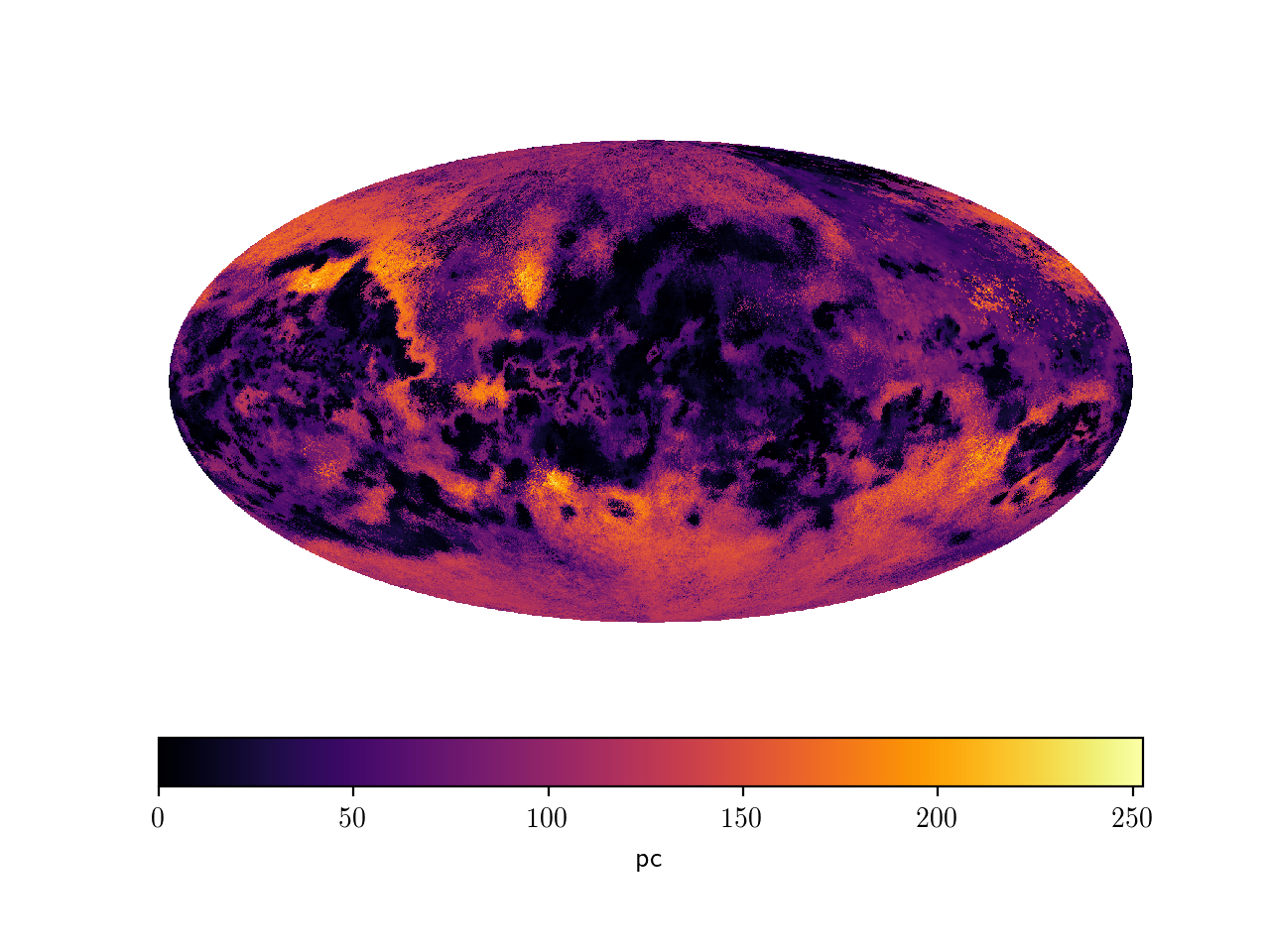}
    \end{subfigure}
        \caption{\label{fig:dist-to-densest}
        A Mollweide projection showing the distance to the voxel with the highest extinction estimate in that direction (left side) and corresponding uncertainty map (right side).
}                               %
\end{figure*}

Using our reconstruction, we can determine distances to nearby dust clouds. %
We derive two distance maps. %
Figure\,\ref{fig:dist-to-nearest} shows the distance to the nearest dust clouds in all directions, as well as an uncertainty on that distance estimate. %
We note that we computed the distance by checking for the first voxel that exceeds a the threshold of $0.005$ e-folds per pc of extinction density. %
Some of our samples do not exclude the existence of nearby dense clouds, which raises the uncertainty in the corresponding directions tremendously. %
Figure\,\ref{fig:dist-to-densest} shows the distance to the densest dust clouds in all directions, as well as an uncertainty on that distance estimate. %
We note that the uncertainty estimate is quite high on the boundaries of dust clouds, as the reconstruction is uncertain which voxel is densest along these lines of sight. %


\subsection{Comparison}
% Here we discuss relations to other papers
\label{sec:comparison}

\begin{figure*}[p]
        \centering
        \begin{subfigure}[t]{.46\textwidth}
                \includegraphics[trim={0.8cm .2cm 2.0cm 1cm}, clip, width=.95\textwidth]{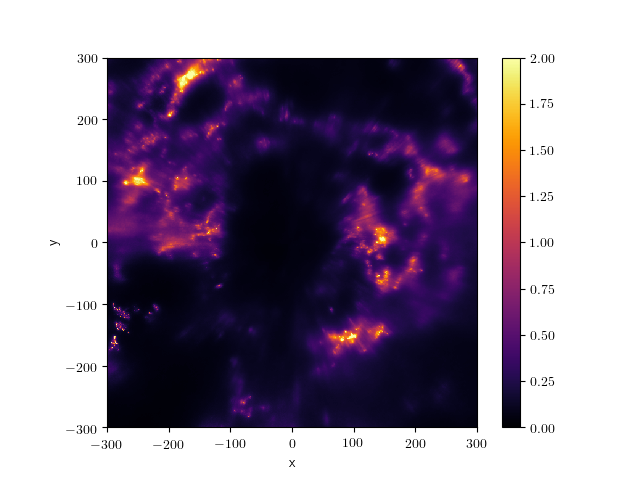}
        \caption{
        }
        \end{subfigure}
        ~
        \begin{subfigure}[t]{.46\textwidth}
                \includegraphics[trim={0.8cm .2cm 2.0cm 1cm}, clip, width=.95\textwidth]{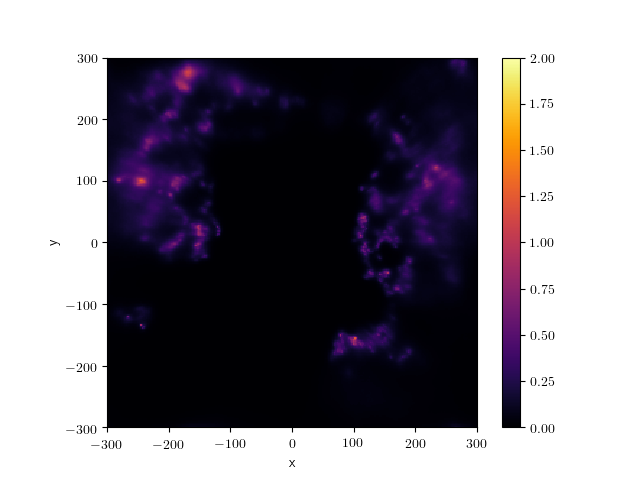}
                \caption{
}
        \end{subfigure}
        \\
        \begin{subfigure}[t]{.46\textwidth}
                \includegraphics[trim={.4cm .2cm 2.0cm 1cm}, clip, width=.95\textwidth]{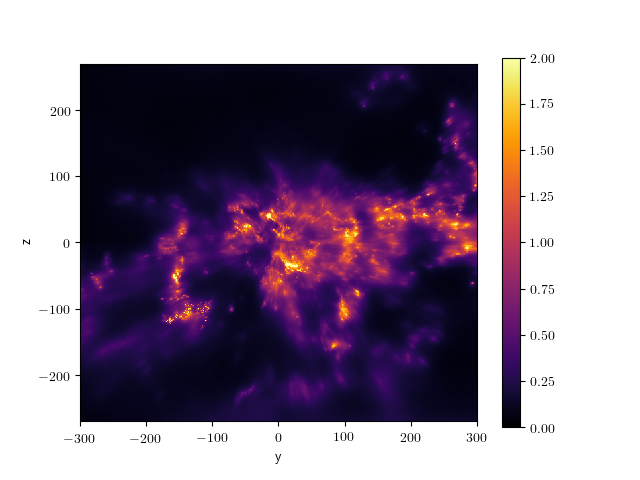}
        \caption{
        }
        \end{subfigure}
        ~
        \begin{subfigure}[t]{.46\textwidth}
                \includegraphics[trim={0.8cm .2cm 2.0cm 1cm}, clip, width=.95\textwidth]{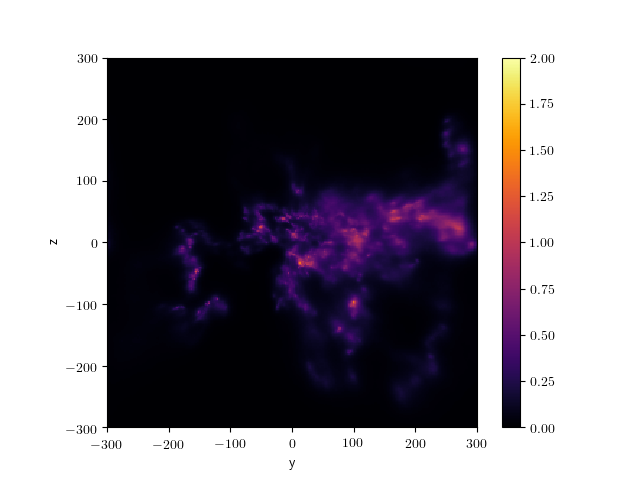}
        \caption{
                }
        \end{subfigure}
        \\
        \begin{subfigure}[t]{.46\textwidth}
                \includegraphics[trim={0.4cm .2cm 2.0cm 1cm}, clip, width=.95\textwidth]{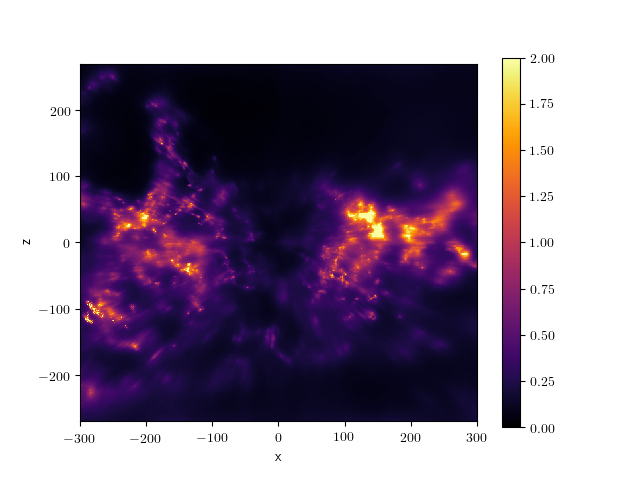}
        \caption{
        }
        \end{subfigure}
        ~
        \begin{subfigure}[t]{.46\textwidth}
                \includegraphics[trim={0.8cm .2cm 2.0cm 1cm}, clip, width=.95\textwidth]{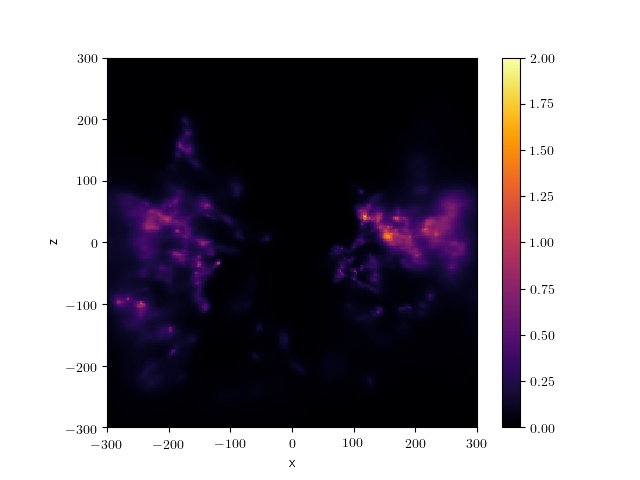}
        \caption{
        }
        \end{subfigure}
        \caption{
        \label{fig:new-old-plane}
        Comparison of column densities of our current reconstruction (left column) and \citet{leike2019charting} (right column).
        The rows show integrated dust extinction for sight lines parallel to the $z$-, $x$-, and $y$-axes, respectively.
        }                       %
\end{figure*}

\begin{figure*}[p]
        \centering
        \begin{subfigure}[t]{.46\textwidth}
                \includegraphics[trim={0.8cm .2cm 2.0cm 1cm}, clip, width=.95\textwidth]{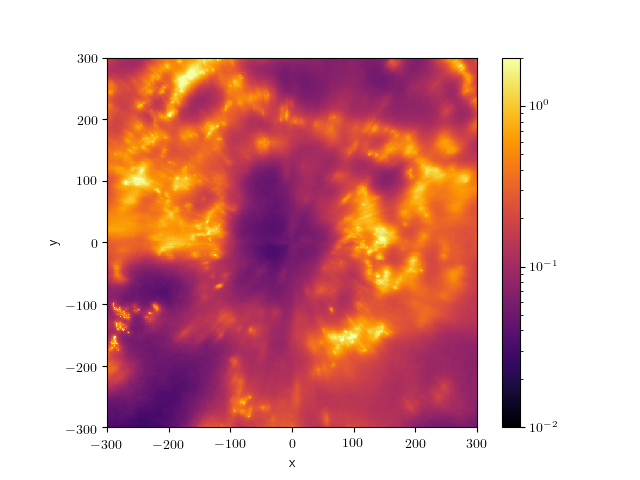}
        \caption{
        }
        \end{subfigure}
        ~
        \begin{subfigure}[t]{.46\textwidth}
                \includegraphics[trim={0.8cm .2cm 2.0cm 1cm}, clip, width=.95\textwidth]{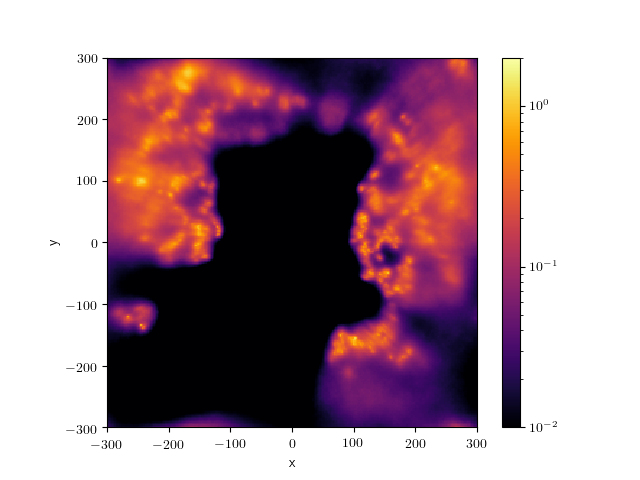}
                \caption{
}
        \end{subfigure}
        \\
        \begin{subfigure}[t]{.46\textwidth}
                \includegraphics[trim={0.4cm .2cm 2.0cm 1cm}, clip, width=.95\textwidth]{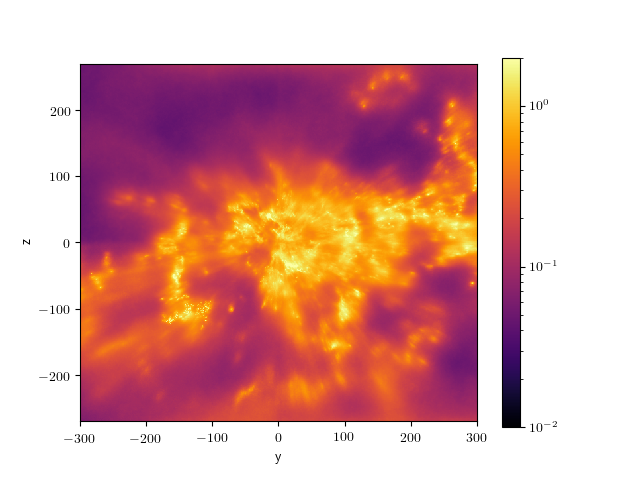}
        \caption{
        }
        \end{subfigure}
        ~
        \begin{subfigure}[t]{.46\textwidth}
                \includegraphics[trim={0.8cm .2cm 2.0cm 1cm}, clip, width=.95\textwidth]{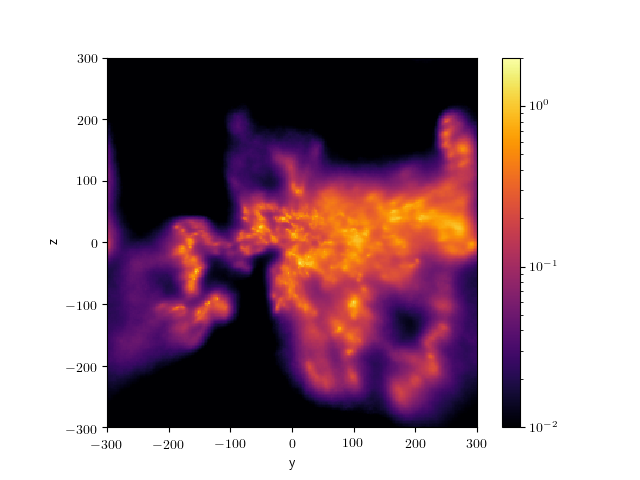}
        \caption{
                }
        \end{subfigure}
        \\
        \begin{subfigure}[t]{.46\textwidth}
                \includegraphics[trim={0.4cm .2cm 2.0cm 1cm}, clip, width=.95\textwidth]{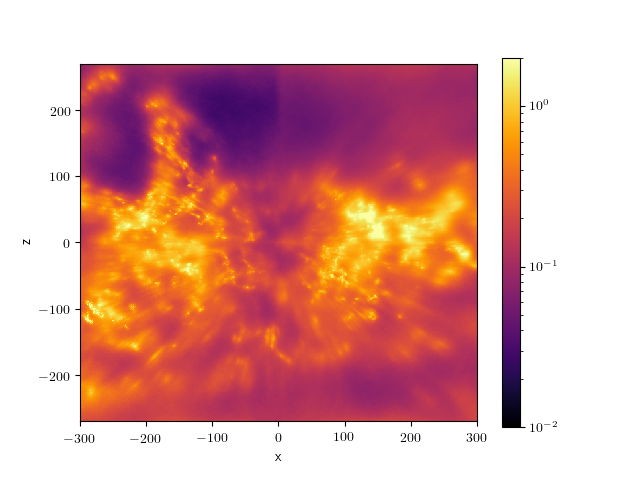}
        \caption{
        }
        \end{subfigure}
        ~
        \begin{subfigure}[t]{.46\textwidth}
                \includegraphics[trim={0.8cm .2cm 2.0cm 1cm}, clip, width=.95\textwidth]{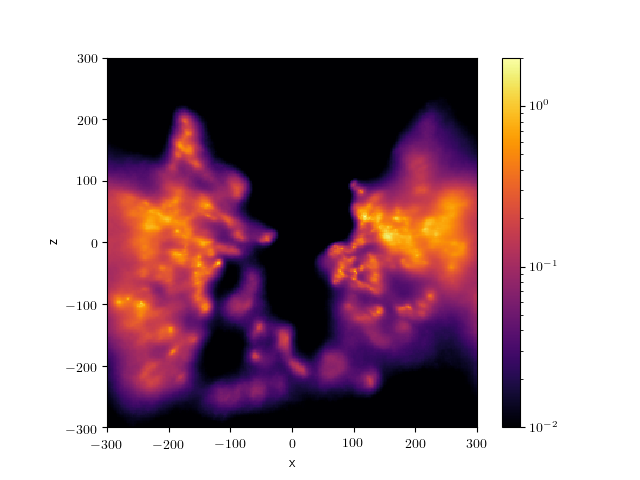}
        \caption{
        }
        \end{subfigure}
        \caption{
        \label{fig:new-old-logplane}
        As Fig.\,\ref{fig:new-old-plane} but on logarithmic scale.
        }                       %
\end{figure*}

An implicit assumption of the algorithm is that the voxels are smaller than the achievable resolution. %
Phrased in physical terms, an increase in pixel resolution can be regarded as a renormalization, and we need to reach the continuous limit, meaning the limit of negligible discretization effects, for the algorithm to work. %
This is a byproduct of the inference of the power spectrum, if the achieved posterior resolution is of the order of the imposed voxel resolution, then the reconstruction changes drastically from one voxel to another, and the extinction of sources behind an affected voxel also changes dramatically at the boundary. %
This sudden change in extinction is not compatible with a falling spatial correlation power law in Fourier space, thus the reconstructed power will fall less steeply than the real one. This would significantly hamper the ability of the algorithm to extrapolate between measurements. %
We avoid this behavior by significantly increasing resolution compared to our previous reconstruction \cite{leike2019charting}. %
However, it is conceivable that the reconstruction would still benefit from increasing the amount of voxels. %
We recommend distrusting the smallest scales of our reconstruction, only at scales of $2\,\text{pc}$ or higher can the result be considered to be stable. %
This resolution limit was deduced from the reconstructed logarithmic correlation kernels as seen in Fig. \ref{fig:kernels}. %
At this limit, the reconstructed correlation kernels of the logarithmic dust extinction density have fallen to $10\%$. A comparison of our results to \citet{leike2019charting} can be found in Fig.\,\ref{fig:new-old-plane}; see Fig.\,\ref{fig:new-old-logplane} for a logarithmic version. %

\begin{figure*}[p]
        \centering
        \begin{subfigure}[t]{.46\textwidth}
                \includegraphics[trim={0.8cm .2cm 2.0cm 1cm}, clip, width=.95\textwidth]{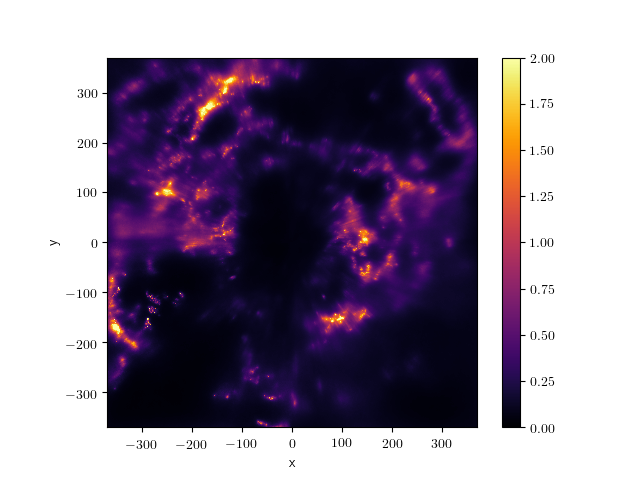}
        \caption{
        }
        \end{subfigure}
        ~
        \begin{subfigure}[t]{.46\textwidth}
                \includegraphics[trim={0.8cm .2cm 2.0cm 1cm}, clip, width=.95\textwidth]{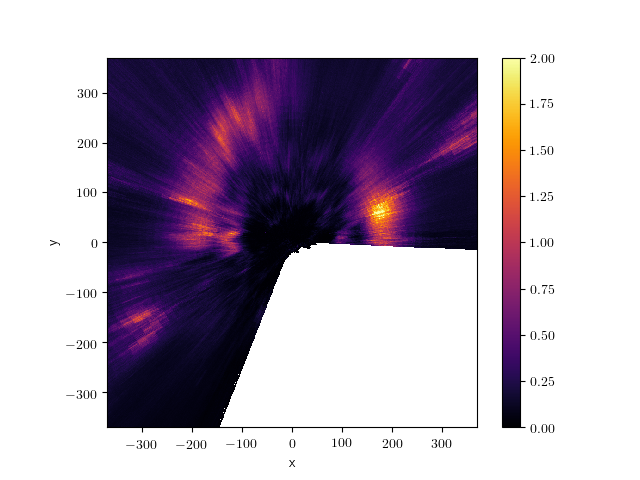}
                \caption{
}
        \end{subfigure}
        \\
        \begin{subfigure}[t]{.46\textwidth}
                \includegraphics[trim={0.4cm .2cm 2.0cm 1cm}, clip, width=.95\textwidth]{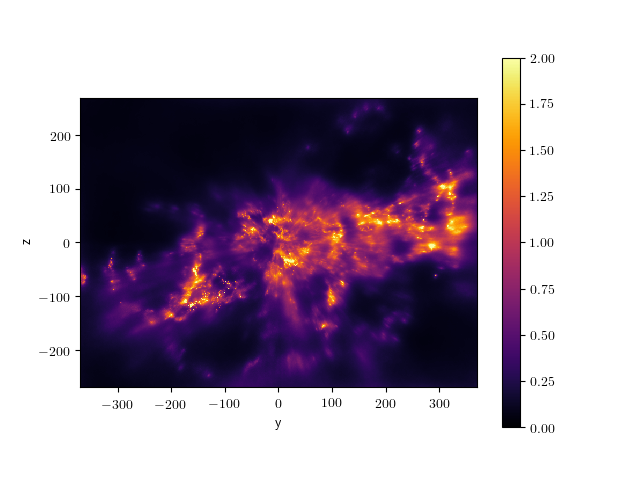}
        \caption{
        }
        \end{subfigure}
        ~
        \begin{subfigure}[t]{.46\textwidth}
                \includegraphics[trim={0.4cm .2cm 2.0cm 1cm}, clip, width=.95\textwidth]{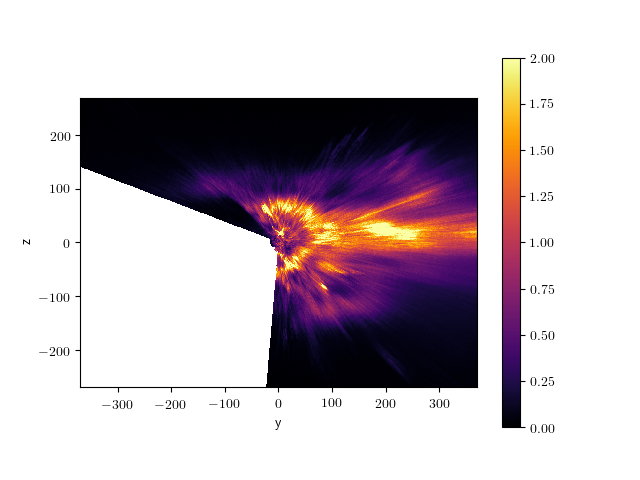}
        \caption{
                }
        \end{subfigure}
        \begin{subfigure}[t]{.46\textwidth}
                \includegraphics[trim={0.4cm .2cm 2.0cm 1cm}, clip, width=.95\textwidth]{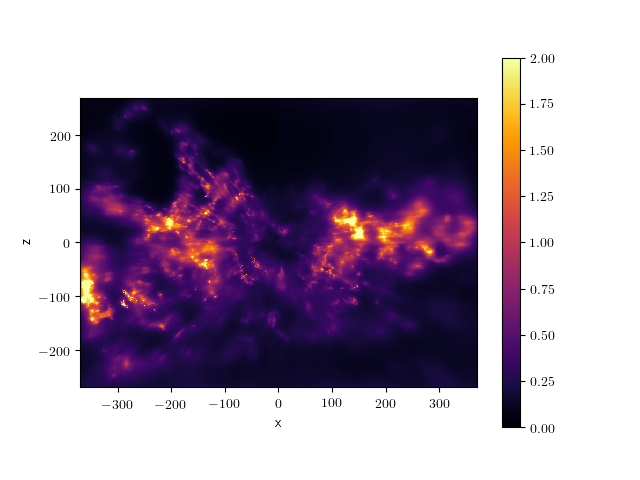}
        \caption{
        }
        \end{subfigure}
        ~
        \begin{subfigure}[t]{.46\textwidth}
                \includegraphics[trim={0.4cm .2cm 2.0cm 1cm}, clip, width=.95\textwidth]{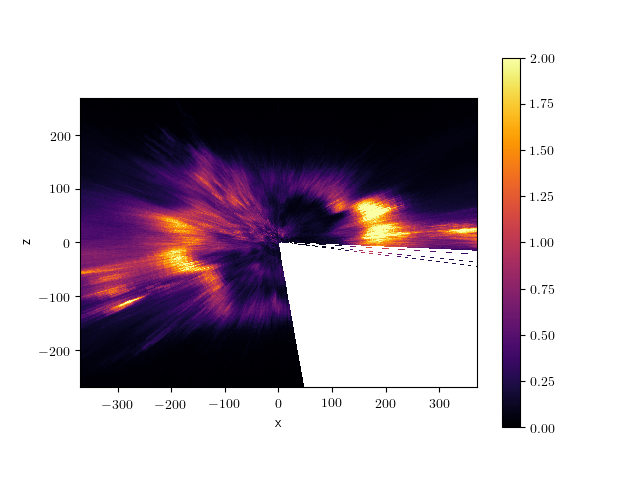}
        \caption{
        }
        \end{subfigure}
        \caption{
        \label{fig:new-green-plane}
        Column density comparison of our current reconstruction (left column) and that of \citet{green20193d} (right column).
        The rows show integrated dust extinction for sightlines parallel to the $z$-, $x$-, and $y$-axes, respectively.
        We note that for \citet{green20193d}, we show the integrated extinction only if more than $50\%$ of the projected voxels exist in the reconstruction. %
        }           %
\end{figure*}

\begin{figure*}[p]
        \centering
        \begin{subfigure}[t]{.46\textwidth}
                \includegraphics[trim={0.8cm .2cm 2.0cm 1cm}, clip, width=.95\textwidth]{our_logplanemean2.png}
        \caption{
        }
        \end{subfigure}
        ~
        \begin{subfigure}[t]{.46\textwidth}
                \includegraphics[trim={0.8cm .2cm 2.0cm 1cm}, clip, width=.95\textwidth]{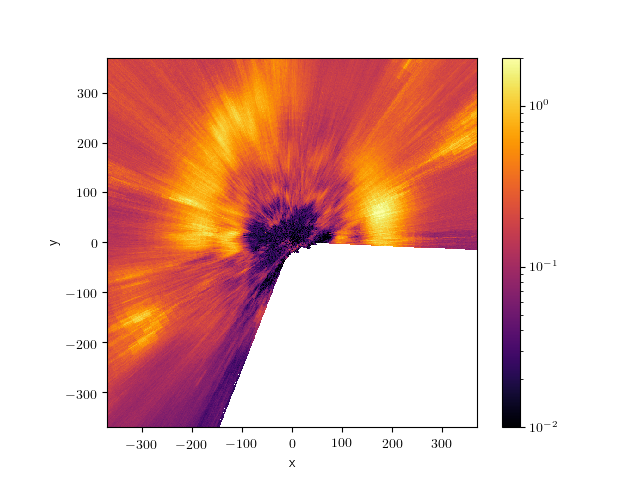}
                \caption{
}
        \end{subfigure}
        \\
        \begin{subfigure}[t]{.46\textwidth}
                \includegraphics[trim={0.4cm .2cm 2.0cm 1cm}, clip, width=.95\textwidth]{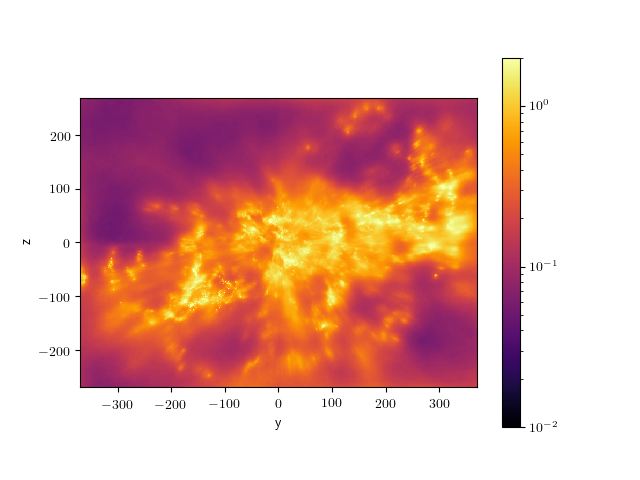}
        \caption{
        }
        \end{subfigure}
        ~
        \begin{subfigure}[t]{.46\textwidth}
                \includegraphics[trim={0.4cm .2cm 2.0cm 1cm}, clip, width=.95\textwidth]{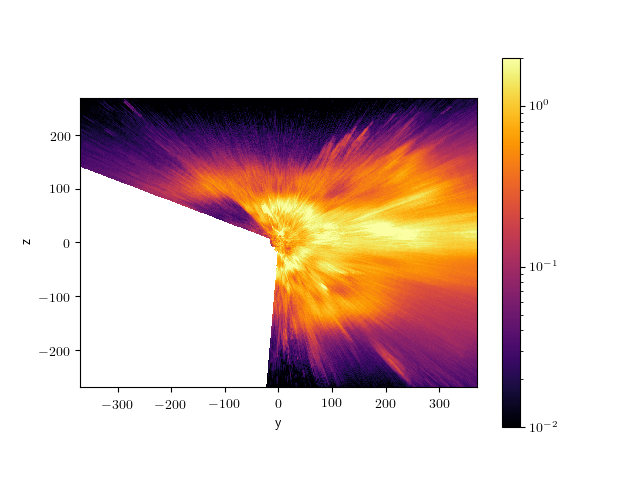}
        \caption{
                }
        \end{subfigure}
        \\
        \begin{subfigure}[t]{.46\textwidth}
                \includegraphics[trim={0.4cm .2cm 2.0cm 1cm}, clip, width=.95\textwidth]{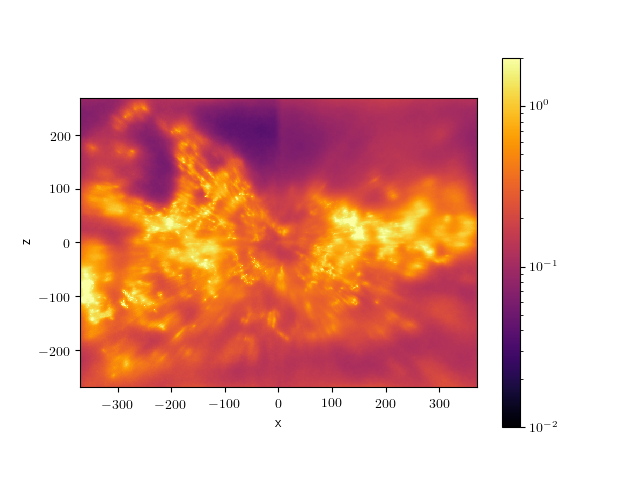}
        \caption{
        }
        \end{subfigure}
        ~
        \begin{subfigure}[t]{.46\textwidth}
                \includegraphics[trim={0.4cm .2cm 2.0cm 1cm}, clip, width=.95\textwidth]{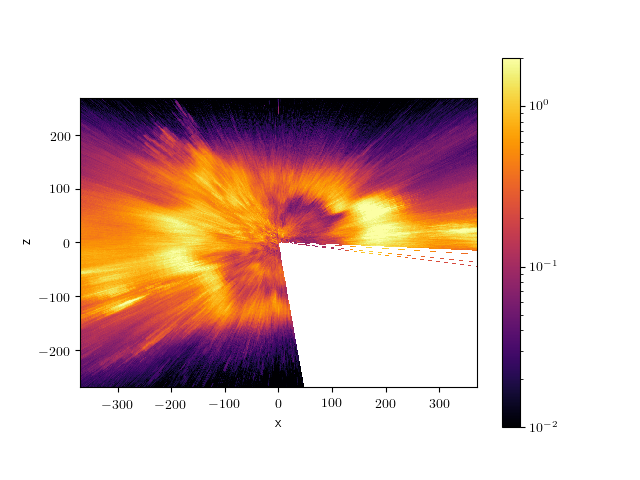}
        \caption{
        }
        \end{subfigure}
    \caption{\label{fig:new-green-logplane}
        As Fig.\,\ref{fig:new-green-plane} but on logarithmic scale.
        }                       %
\end{figure*}

\begin{figure*}[p]
        \centering
        \begin{subfigure}[t]{.46\textwidth}
                \includegraphics[trim={0.8cm .2cm .5cm .5cm}, clip, width=.9\textwidth]{our_loghealpixmean.png}
        \caption{
        }
        \end{subfigure}
        ~
        \begin{subfigure}[t]{.46\textwidth}
                \includegraphics[trim={0.8cm .2cm .5cm .5cm}, clip, width=.9\textwidth]{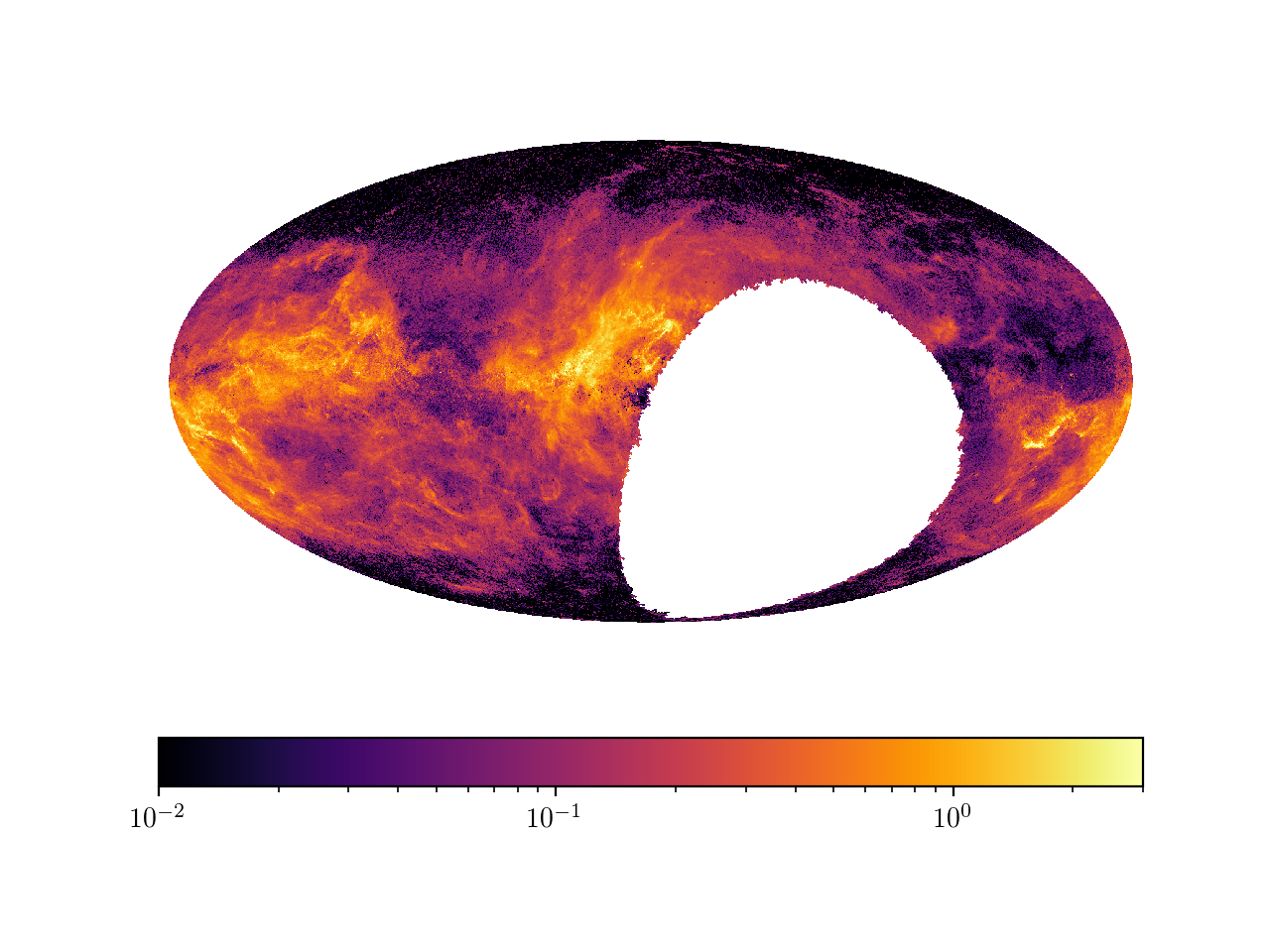}
                \caption{
}
        \end{subfigure}
        \\
        \begin{subfigure}[t]{.46\textwidth}
                \includegraphics[trim={0.8cm .2cm 2.0cm 1cm}, clip, width=.95\textwidth]{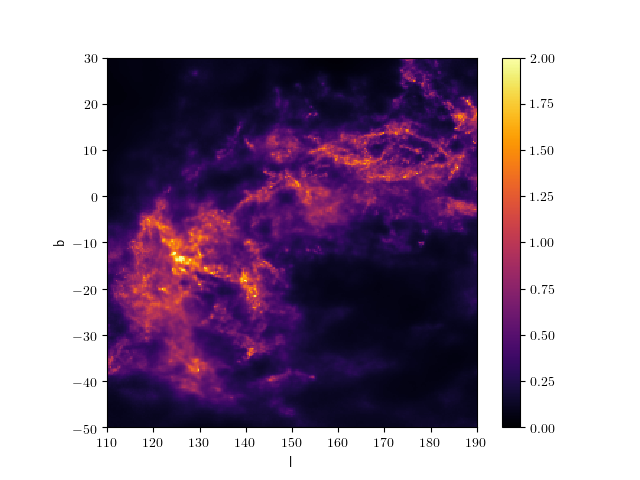}
        \caption{
        }
        \end{subfigure}
        ~
        \begin{subfigure}[t]{.46\textwidth}
                \includegraphics[trim={0.8cm .2cm 2.0cm 1cm}, clip, width=.95\textwidth]{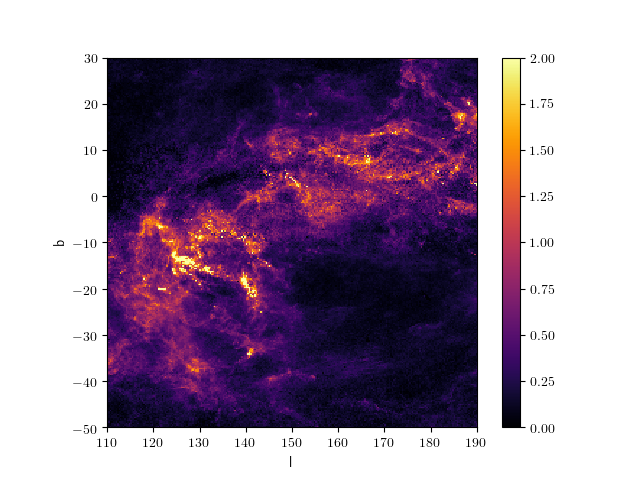}
        \caption{
                }
        \end{subfigure}
        \\
        \begin{subfigure}[t]{.46\textwidth}
                \includegraphics[trim={0.8cm .2cm 2.0cm 1cm}, clip, width=.95\textwidth]{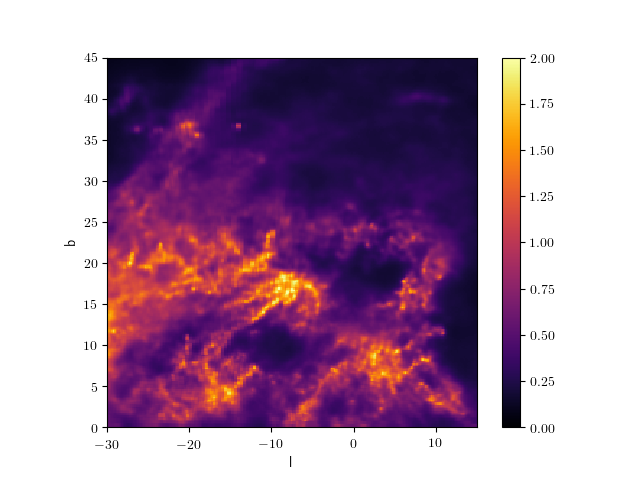}
        \caption{
        }
        \end{subfigure}
        ~
        \begin{subfigure}[t]{.46\textwidth}
                \includegraphics[trim={0.8cm .2cm 2.0cm 1cm}, clip, width=.95\textwidth]{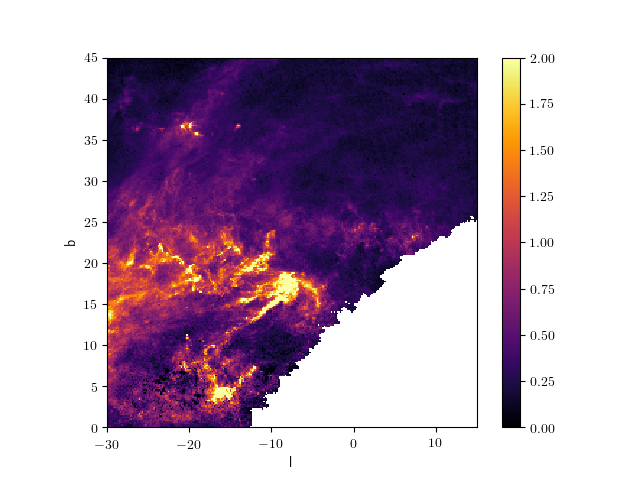}
        \caption{
        }
        \end{subfigure}
    \caption{\label{fig:sky-comparison}
        Comparison of integrated extinction of our reconstruction (left column) and that of \citet{green20193d} (right column) in sky projection.
    The rows show integrated dust extinction out to the boundary of our $740\,\text{pc}\times740\,\text{pc}\times540\,\text{pc}$ box in an all sky view (first row), as well as two selected directions towards the Galactic anti-center (middle row) and center (last row).
        }
\end{figure*}

We compare our results to the map of \citet{green20193d}. %
Figure\,\ref{fig:new-green-plane} shows column density comparisons of the two reconstructions. %
Figure\,\ref{fig:new-green-logplane} shows the same column densities, but on a logarithmic scale. %
A more detailed comparison to \citet{green20193d} in angular coordinates can be seen in Fig.\,\ref{fig:sky-comparison}.

In contrast to our old map, we used the dataset of \cite{anders2019photo}, which provides more sources and tighter constraints on the parallax and $G$-band extinction than the previously used Gaia data. %
The new reconstruction has a volume of $800\,\text{pc}\times800\,\text{pc}\times\,600\text{pc}$, compared to the $(600\,\text{pc})^3$ cube in \citet{leike2019charting}. %
Furthermore, using a designated fortran routine for the computation of the line-of-sight integrals lead to the necessary speedup to handle the additional data constraints and significantly more degrees of freedom. %
Finally, in the new reconstruction, the parallax error is propagated into the measurement error, causing extinction values with stars of high parallax error to be less informative. %
In Fig.\,\ref{fig:new-old-plane}, one can see dust column densities along Galactic $x$, $y$, and $z$ coordinates. %
Both dust maps agree on the morphology of large dust clouds on large scales. %
However, the current dust map contains significantly more dust. %
Part of the reason is that the data we use in the reconstruction of this paper has higher resolution and lower noise, allowing more dust to be reconstructed. %
We also believe the data used in \citet{leike2019charting} to be slightly biased toward underestimating the amount of dust, an effect that accumulates in a reconstruction that uses many data points. %
In contrast, the data used in this reconstruction might have a tendency to overestimate the amount of dust, despite our efforts to calibrate the zero point (see Sect.\,\ref{sec:noise-statistic}). %

\begin{figure}
                \includegraphics[trim={0.6cm .2cm .5cm 0.5cm}, clip, width=.45\textwidth]{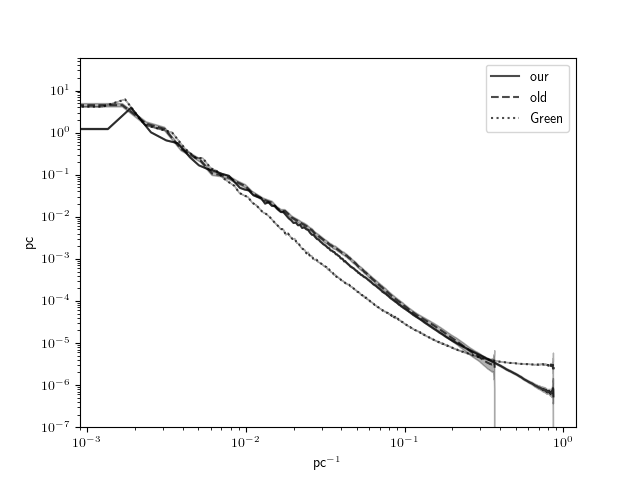}
        \caption{\label{fig:power-spectra}
        Empirical power spectra of the dust extinction density of this paper (solid line), \citet{leike2019charting} (dashed line) and the reconstruction of \citet{green20193d} (dotted line).} %
\end{figure}

\begin{figure}
                \includegraphics[trim={0.6cm .2cm .5cm .5cm}, clip, width=.45\textwidth]{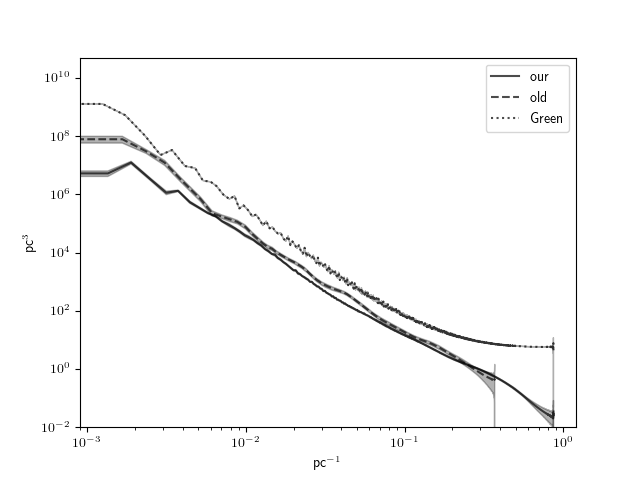}
        \caption{\label{fig:logpower-spectra}
        Empirical power spectra of the logarithmic dust extinction density of this paper (solid line), \citet{leike2019charting} (dashed line) and the reconstruction of \citet{green20193d} (dotted line).} %
\end{figure}

Furthermore, we reconstructed our correlation kernel nonparametrically, which should lead to an unbiased estimate of the power spectrum.
Figure\,\ref{fig:power-spectra} shows power spectra of \citet{leike2019charting}, the reconstruction of this paper, and of the reconstruction of \citet{green20193d}.
Our new reconstruction and \citet{leike2019charting} seem to have quite consistent power spectra. %
The general tendency of the falling power law is also remarkably consistent with \citet{green20193d}, however at scales of a few parsec the power spectrum of \citet{green20193d} flattens, which we believe to be an artifact of how we put their reconstruction on a cartesian grid. For example, the boundaries of the reconstructions intrinsic voxels introduce steep cuts that flatten the resulting power spectrum. %
However, none of the power spectra are consistent within the uncertainty estimates. %
While this seems problematic, one has to bear in mind that all reconstructions focus on dust in differing regions, potentially explaining the difference in the power spectrum. %
In Fig.\,\ref{fig:logpower-spectra}, we show the power spectra of the logarithmic reconstructions.
These seem to be less consistent in general, however one has to bear in mind that the logarithmic power spectrum is dominated by regions of low dust content, as these occur more frequently. %
Using Gaia data, our method was found to underestimate low dust regions, and we anticipated that with the starhorse data, we would tend to overestimate low dust regions. %
Nonetheless, we find that the spectral index of $2.82\pm0.022$ at scales from $2.3\,\text{pc}$ to $125\,\text{pc}$ is compatible with the empirical spectral index of \citet{leike2019charting} within a $2\sigma$ joint uncertainty margin. %
The spectral index of the empirical power spectrum of \citet{leike2019charting} is $3.2\pm0.14$.\footnote{We note that \cite{leike2019charting} reported a spectral index of $3.1$ for the reconstructed power spectrum. For this paper, we instead chose to analyze the power spectrum of the resulting maps, which yields slightly different values but enables us to derive uncertainty estimates for all compared maps in the same way. }%
The logarithmic power spectrum of \citet{green20193d} seems to be inconsistent with our measurements. %
However, this effect is probably due to how we treat the missing values in that map, where a quarter of the sky was not measured. %
We have to set these values and every possible choice will impact the derived power spectra. %
We chose to set them to $10^{-7}$, which has minimal impact on the power spectrum on a linear scale, but biases the power spectrum of the logarithmic dust extinction density and could potentially explain the difference. %

\begin{figure}
                \includegraphics[trim={0.8cm .2cm .5cm 0.5cm}, clip, width=.45\textwidth]{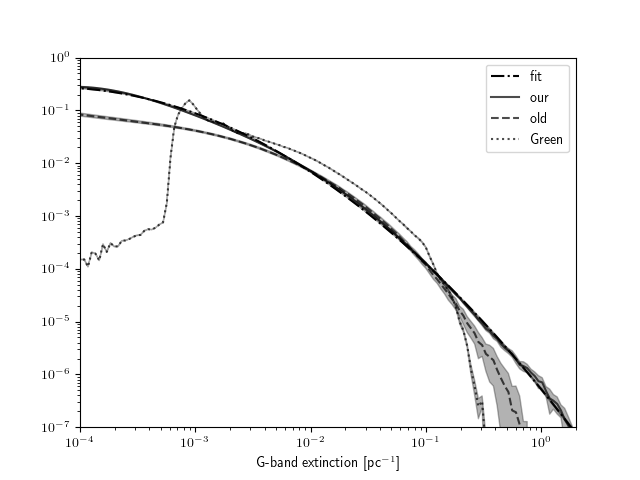}
        \caption{\label{fig:loghistogram}
        Histogram of dust extinction density per voxel of this paper (solid line), \citet{leike2019charting} (dashed line) and the reconstruction of \citet{green20193d} (dotted line).
        We overplot a log-normal model that was fit to our reconstructed logarithmic extinction density (dash-dotted line). %
        The curve of the fit is described by $f(x) \propto \text{exp}\left(-0.5\sigma^{-2}\left(\text{ln}(x)-m\right)\right)$ with $\sigma = 1.906$ and $m = -9.79$ and follows our empirical distribution function closely.} %
\end{figure}

Figure\,\ref{fig:loghistogram} shows a histogram of dust extinction density per voxel. %
One can see a good agreement between the histogram of our old and our current reconstruction in the region between $10^{-3}\,\text{pc}^{-1}$ and $10^{-1}\,\text{pc}^{-1}$. %
A dust extinction density of $10^{-4}\,\text{pc}^{-1}$ integrated to the boundary of our simulation cube yields an integrated extinction of $0.046$, which is below our noise level even when pooling the information of many stars. %
For this reason, we do not show the histogram below $10^{-4}\,\text{pc}^{-1}$ , as its shape is mostly dependent on how the reconstruction extrapolates into a dustless region. % 
From the histogram, it can be seen that the dust density is well described by a log-normal distribution. %
We note that since we only show the part of the histogram that has high signal to noise, this result should be relatively unbiased by our choice of prior. %
The fit log-normal model has a standard deviation of $\sigma=1.906\pm0.009$ and a mean of $m=-9.79\pm0.04$. %

\subsection{Using the reconstruction}
\label{sec:using-the-reconstruction}

One should note that the reconstruction shows a non-negligible amount of dust in the local bubble. %
We believe that the level found is an artifact of our noise statistic. %
As described in Sect.\,\ref{sec:likelihood}, our data model involves some heuristics that might systematically affect the reconstruction. %
This causes estimates made with this data to be biased, and we were not able to fully correct for this bias. %
When integrating the reconstructed dust density to $70\,\text{pc}$, we find that the nearby dust looks like a smeared out version of farther dust clouds, indicating that it is indeed an artifact related to systematic data biases. %

The posterior samples of the extinction density are available for download\footnote{\url{https://doi.org/10.5281/zenodo.3750926}, DOI 10.5281/zenodo.3750926 or at the CDS \url{http://cds.u-strasbg.fr/}.}. % TODO INSERT LINK TO CDS
When using the reconstruction, we advise that you beware of systematic overestimations of dust, especially in the local bubble.
When deriving numeric quantities, we advise doing so for every sample and then estimating the mean and standard deviation of the results in order to get an error estimation. %

\subsection{Implications}

Our map can be used to constrain simulations of the ISM. %
For example, in simulations of radiatively cooling dust clouds in hot winds, it has been shown that dust density power spectra are flatter than was previously thought \citep{sparre2019physics}. %
Our maps show power spectra compatible with these simulations, and morphologically similar structures. %
Our reconstructed spectral index of $2.82\pm0.022$ at scales from $2.3\,\text{pc}$ to $125\,\text{pc}$ could be used to constrain parameters of sub-grid models of simulations of the ISM. %
Furthermore, we find the density histogram of the logarithm of the $G$-band dust extinction density in e-folds per parsec shown in Fig.\,\ref{fig:loghistogram} is well described by a log-normal distribution with standard deviation $\sigma=1.906\pm0.009$ and mean $m=-9.79\pm0.04$ on extinction density scales from $10^{-4}\,\text{pc}^{-1}$ to $1\,\text{pc}^{-1}$. %

\section{Conclusion}
\label{sec:conclusion}

We were able to reconstruct the dust clouds within $\sim 400\,\text{pc}$ of the Sun down to a resolution of $2\,\text{pc}$, improving in resolution and volume on our previous reconstruction \citep{leike2019charting}. %
The resulting map is public and can be downloaded; see Sect.\,\ref{sec:using-the-reconstruction} for details. %
Distances to and densities of all dust clouds larger than $2\,\text{pc}$ are expected to be well constrained by the reconstruction. %
We report our estimate on the power spectrum of the dust extinction density as well as the logarithmic density. %
Furthermore, we provide a histogram of dust densities in the interstellar medium and find them to be well described by a log-normal model. %
We hope that our diverse summary statistics allow simulations of the ISM to be constrained. %

%______________________________________________________________

\begin{acknowledgements}
        We acknowledge fruitful discussions with S. Hutschenreuter, J. Knollm\"uller, P. Arras, A. Kostic, and others from the information field theory group at the MPI for Astrophysics, Garching.
We acknowledge the support by the DFG Cluster of Excellence "Origin and Structure of the Universe". The prototypes for the reconstructions were carried out on the computing facilities of the Computational Center for Particle and Astrophysics (C2PAP). %
    We acknowledge support by the Max-Planck Computing and Data Facility (MPCDF). The main computation for the reconstructions were carried on compute cluster FREYA. %

This work has made use of data from the European Space Agency (ESA) mission
{\it Gaia} (\url{https://www.cosmos.esa.int/gaia}), processed by the {\it Gaia}
Data Processing and Analysis Consortium (DPAC,
\url{https://www.cosmos.esa.int/web/gaia/dpac/consortium}). Funding for the DPAC
has been provided by national institutions, in particular the institutions
participating in the {\it Gaia} Multilateral Agreement.
\end{acknowledgements}

%-------------------------------------------------------------------

\bibliography{ift}
\bibliographystyle{aa}
\appendix
\section{interpolation scheme}
\label{sec:interpolation-scheme}

To parallelize the reconstruction, we reconstructed the eight octants of the coordinate system independently, with a $20\,\text{pc}$ overlap region. %
To get one final reconstruction, we have to glue these reconstructions together and specify how we deal with the overlap region. %
We do so using a differentiable, variance-preserving interpolation scheme, meaning that if the octants are differentiable then the result is differentiable; and the final samples have at least the variance that the individual reconstructions imposed. %
We compute the uncorrected interpolated logarithmic extinction samples $\tau^\prime(x)_j$ from the samples of the eight octants $o^i(x)_j$ as
\begin{align}
    \tau^\prime(x)_j &= \sum_i w_i(x) o^i(x)_j \ .
\end{align}
Thus, the weights $w_i(x)$ can be computed as %
\begin{align}
    w_i(x) &= \prod_{k=0}^2\left|b_k(i)-f\left(\frac{x_k-9\,\text{pc}}{18\,\text{pc}}\right)\right|\ ,\\
    \text{where } f(x) &=
    \begin{cases}
        0 & x \in (-\infty,0]\\
        x^2(3-2x) & x\in(0,1)\\
        1 & \text{ otherwise,}
    \end{cases}
\end{align}
and $b_k(i)$ denotes the $k$-th digit of $i$ in binary format. %
Noteworthy properties of this scheme are that the weights sum to one
\begin{align}
    \forall x\,\sum_i w_i(x) = 1,
\end{align}
and the polynomial $f$ is the unique polynomial of degree $3$ so that
\begin{align}
    f(0) = 0\ ,\\
    f(1) = 1\ ,\\
    \frac{\partial f}{\partial x}(0) = 0\ ,\\
    \frac{\partial f}{\partial x}(1) = 0 \ .
\end{align}
By using this interpolation scheme, only voxels that have a coordinate $x_k$ of which the absolute value is at most $8\,\text{pc}$ get nonzero contributions from more than one octant. In other words, %
we cut away the outermost $2\,\text{pc}$ of all reconstructions, which mitigates artifacts from periodic boundary conditions. %
From the preliminary logarithmic extinction density $\tau^\prime(x)_j,$ we can compute the logarithmic sample mean $\bar{\tau}(x)$ as %
\begin{align}
    \bar{\tau} &= \frac{1}{N}\sum_j \tau^{\prime}_j\ .
\end{align}
Here, $N$ denotes the number of samples. %
The variance of $\tau^{\prime}_j$ is artificially low at overlapping regions, as independent samples were averaged. %
We correct for this effect and compute the overall logarithmic extinction density samples $\tau(x)_{j}$ as %
\begin{align}
    \tau(x)_{j} &= \frac{\tau^\prime(x)_j-\bar{\tau}(x)}{\sqrt{\sum_i w_i(x)^2}}+\bar{\tau}(x) \ .
\end{align}

\end{document}